\documentclass[12pt]{article}
\usepackage{amsmath, amsfonts, amssymb}
\begin{document}
\newtheorem{thm}{Theorem}
\newtheorem{lemma}{Lemma}[section]
\newtheorem{prop}[lemma]{Proposition}
\newtheorem{cor}[lemma]{Corollary}
\newtheorem{fact}{Fact}
\def\endproof{\hspace{5.25in}\rule{3mm}{3mm}}
\def\endrmk{\hspace{5.25in}\rule{2mm}{2mm}}
\def\endpf{\quad\rule{2mm}{2mm}}
\allowdisplaybreaks

\begin{center}\textbf{On the Asymptotics for the Vacuum}\\ \textbf{Einstein
Constraint Equations}\\

\bigskip

\begin{tabular}{cc}

Justin Corvino & Richard M. Schoen \\

Brown University & Stanford University \\

\small{ \textrm{corvino@math.brown.edu}} & \small\textrm{schoen@math.stanford.edu}\\



\end{tabular}

\end{center}

\begin{abstract}
Given asymptotically flat initial
data on $M^3$ for the vacuum Einstein field equation, and given a bounded domain in $M$, we construct
solutions of the vacuum constraint equations which agree with the
original data inside the given domain, and are identical to that of a suitable
Kerr slice (or identical to a member of some other admissible family of
solutions) outside a large ball in a given end.  The data for which this construction works is shown to be dense in an appropriate topology on the space of asymptotically flat solutions of the vacuum constraints.  This construction
generalizes work in \cite{cor:schw}, where the time-symmetric case was
studied.    
\end{abstract}

\section{Introduction.}

We study the asymptotics of solutions to the vacuum constraint equations. 
In particular, we show how to approximate given initial data for the Einstein
vacuum equation by data which is exactly that of a spacelike slice of a
suitably chosen Kerr metric outside a compact set.  For the time-symmetric
case, the constraints reduce to the equation
$R(g)=0$, and it was shown in \cite{cor:schw} how a Schwarzschild exterior can
be selected and glued to asymptotically flat (AF) time-symmetric data and
\textit{compactly} perturbed to a solution of the time-symmetric constraint.  In
the non-time-symmetric case, one has to consider the second fundamental form, and
hence linear and angular momentum of the data at infinity, for which the Kerr
solution (for example) is aptly suited.  We note that these constructions show that
unique continuation fails in a serious way, but it also points to the reality that a
compact piece of the data determines (in some sense) only part of the asymptotic
structure, and indeed that which it determines is only the part that is governed by
the energy-momentum and angular momentum. Having vacuum initial data which is
identical to Kerr data outside a compact set is important for understanding the
global evolution for the Einstein equations. In particular, such data evolves to produce a spacetime with particularly nice behavior at null infinity.\\

We briefly recall the basic setup of the constraint equations. A solution of the
vacuum Einstein equation is a Lorentzian manifold $(\mathcal{S}^4,\bar{g})$
satisfying  
\begin{eqnarray*}
\label{eq:Ein}
Ric(\bar{g})-\frac{1}{2}R(\bar{g})\, \bar{g}=0.
\end{eqnarray*}
In this equation, $Ric(\bar{g})$ and $R(\bar{g})$ denote the Ricci and scalar
curvatures, respectively, of the metric $\bar{g}$.  By taking the trace we see that
this is equivalent to $Ric(\bar{g})=0$.  It is well-known that Einstein's equation
admits an initial value formulation, in which the vacuum initial data consist of an
oriented three-manifold $M^3$, a Riemannian metric $g$ and a symmetric $(0,2)$-tensor
$K$ on $M$.  The Gauss and Codazzi equations provide the constraints upon $g$ and $K$
in order that they form, respectively, the induced metric and second fundamental form
of $M$ inside a Ricci-flat spacetime $(\mathcal{S},\bar{g})$; in the vacuum case these
constraints are \cite{wa:gr}
\begin{eqnarray*}
\label{eq:Cons1}
R(g)-|K|^2+H^2 &=&0 \\
div_g(K)-\nabla H &=& 0.
\end{eqnarray*}
Here $H=Tr_g(K)=g^{ij}K_{ij}$ denotes the mean curvature, and all quantities are
computed with respect to $g$.\\

We rewrite these equations by introducing the momentum tensor
\[\pi^{ij}=K^{ij}-Tr_g(K) g^{ij}.\]  We also introduce functions $\mathcal{H}$ and
$\Phi$ by 
\begin{eqnarray*}
\nonumber\mathcal{H}(g,\pi)&=&R(g)+\frac{1}{2}\left( Tr_g\pi \right)
^2-|\pi|^2\\
\Phi(g,\pi)&=&\big(\mathcal{H}(g,\pi),div_g\pi\big).
\end{eqnarray*}
The constraints then take the form $\Phi(g,\pi)=0$. \\

Two well-known solutions of the vacuum Einstein equation are the Schwarzschild
solution and the Kerr solution.  The Schwarzschild metric is characterized by
rotational symmetry and is \textit{static}; indeed outside the horizon we have
coordinates in which the Schwarzschild metric takes the form
\[g_S=-\left(1-\frac{2m}{r}\right)dt^2 +
\left(1-\frac{2m}{r}\right)^{-1}dr^2+r^2\;d\Omega^2.\] In particular we see that the
metric is static, as $\frac{\partial}{\partial t}$ is a timelike Killing field which
is orthogonal to the time-slices.  The Kerr solution is axisymmetric and stationary;
the analogous Killing field is not orthogonal anymore to the time-slices (physically, the
black hole is rotating), and this can be easily seen in Boyer-Lindquist coordinates
\cite{mtw}, \cite{wa:gr}.\\


What is more important for us is that these solutions are actually \textit{families} of
solutions of the Einstein equations.  In fact by thinking of a fixed asymptotically flat
coordinate system near infinity in the spacetime, by varying the total mass and angular
momentum, and considering families of spacelike slices of these metrics, we get a
ten-parameter family of asymptotically flat solutions of the vacuum constraint equations
near infinity in $\mathbb{R}^3$.  The purpose of considering this family is to control the
energy-momentum $(E,\mathbf{P})$ and the angular momentum $\mathbf{J}$ at (spacelike)
infinity, as well as a quantity $\mathbf{C}$ related to the center-of-mass.  In an
asymptotically flat chart these quantities are defined as limits of integrals over euclidean
spheres with surface measure $d\sigma_g$ and outer normal $\nu$ taken with respect to $g$
\begin{eqnarray}
E&=&\frac{1}{16\pi}\lim\limits_{R\rightarrow\infty}\oint\limits_{|\mathbf{x}|=R}
\sum\limits_{i,j}\left( g_{ij,i}-g_{ii,j}\right) \nu^j d\sigma_g\\
P_i&=&\frac{1}{8\pi}\lim\limits_{R\rightarrow\infty}\oint\limits_{|\mathbf{x}|=R}
\sum\limits_{j}\pi_{ij}\nu^j d\sigma_g\\
J_i&=&\frac{1}{8\pi}\lim\limits_{R\rightarrow\infty}
\oint\limits_{|\mathbf{x}|=R}\sum\limits_{j,k}
\pi_{jk}Y^j_i\nu^k d\sigma_g\\
 C^k &=&\lim\limits_{R\rightarrow\infty}\oint\limits_{|\mathbf{x}|=R}
\sum\limits_{i,j} x^k \left( g_{ij,i}-g_{ii,j}\right) \nu^j d\sigma_g - 
\lim\limits_{R\rightarrow\infty}\oint\limits_{|\mathbf{x}|=R}
\sum\limits_{i}\big(g_{ik} \nu^i- g_{ii}\nu^k \big)  d\sigma_g 
 \end{eqnarray}
In the case when $g$ is AF, conformally flat near infinity and has zero scalar
curvature, the quantity $\mathbf{C}$ is proportional to $m\mathbf{c_0}$, where $m$ is
the mass and $\mathbf{c_0}$ is the coordinate translation which makes the
$|x|^{-2}$-terms in the expansion of the conformal factor vanish \cite{cor:schw}.  We
note that it may be useful to shift the quantity $\mathbf{C}$ (for example an initial
shift of coordinates will do this), and also we note that the vector fields
$Y_i$ are the basic rotation fields in $\mathbb{R}^3$, for example
\[Y_1=x^2\frac{\partial}{\partial x^3}-x^3\frac{\partial}{\partial x^2}.\] The
integrand for the angular momentum is essentially the cross product of the position
vector with the momentum tensor, in analogy with the angular momentum of classical
mechanics.\\
     
We can prescribe conditions under which the above asymptotic integrals
converge for \textit{solutions} of the vacuum constraints.  These conditions are
asymptotic even/odd conditions on the metric tensor and second fundamental form of the
initial data; we call this condition (AC).  The conditions are \begin{eqnarray} 
\nonumber g_{ij}(\mathbf{x})=\delta_{ij}+O(|\mathbf{x}|^{-1}) &\quad &
K_{ij}(\mathbf{x})=O(|\mathbf{x}|^{-2})\\g_{ij}(\mathbf{x})-g_{ij}(-\mathbf{x})=O(|\mathbf{x}|^{-2})
& \quad & K_{ij}(\mathbf{x})+K_{ij}(-\mathbf{x})=O(|\mathbf{x}|^{-3}) \\ \nonumber
g_{ij,k}(\mathbf{x})+g_{ij,k}(-\mathbf{x})=O(|\mathbf{x}|^{-3}) &\quad &
K_{ij,k}(\mathbf{x})-K_{ij,k}(-\mathbf{x})=O(|\mathbf{x}|^{-4})\label{eq:RT}\end{eqnarray} 
We require analogous conditions on successive derivatives as needed. It is also
worthwhile to remark that under these asymptotic conditions on solutions of the constraints, the corresponding
boundary integrals above can also be computed with respect the euclidean metric, and the limiting values are the same, a fact we may use without further comment. \\

In Section 3 of this paper we show that the vacuum initial data sets which
satisfy (AC) are dense in a suitable weighted Sobolev space ($g_{ij}-\delta_{ij}\in
W^{2,p}_{-\delta}$  and $\pi\in W^{1,p}_{-1-\delta}$ for $p>3$, $\delta\in (1/2,1)$) in the
set of all vacuum initial data which satisfy the standard decay assumptions (but not the
asymptotic symmetry). This space is strong enough that the total energy and linear
momentum are continuous on the space. (Of course, the angular momentum is not continuous,
nor does it appear to be well-defined on this more general space.) This theorem is an
analogue for the full constraint equations of the theorem of Schoen-Yau \cite{sy5} in the
time-symmetric case ($K=0$). It is however much more subtle to prove because the general
constraint operator is not as well understood as the time-symmetric operator. We show
that it is possible to achieve a very special type of asymptotic behavior which in
particular satisfies (AC). These special asymptotic conditions require that outside a compact set
there exist a positive function $u$ and a vector field
$X$ so that $g_{ij}=u^4\delta_{ij}$ and
$\pi_{ij}=u^2(L_X\delta-div_\delta(X)\delta_{ij})$ where
$L_X\delta$ denotes the Lie derivative of the euclidean metric $\delta$ with respect to 
$X$. It is not difficult to see that such data automatically satisfies (AC). An important
feature of these asymptotic conditions is that the total energy and momenta (both linear
and angular) may be read off the asymptotics, and the behavior of these quantities
directly effects the asymptotic geometry. This type of asymptotic behavior will be
discussed in detail by the second author in \cite{s2}. The proof of the density of
solutions involves taking an arbitrary solution of the vacuum constraint equations and
patching $g$ to the euclidean metric and
$\pi$ to
$0$ in a large annulus. It is then necessary to reimpose the constraints by solving
a PDE for $(u,X)$. The cut-off data may be viewed as an approximate solution, and
one may attempt to use the inverse function theorem  to correct it to a solution.
The problem which arises is that the corresponding linearization is not clearly an
isomorphism in the natural spaces. This problem is overcome by allowing the extra
flexibility of addition of a suitably chosen family of compactly supported
deformations (of $g,\pi$). In order to make this work, we need to use the fact that
the constraint operator (at an arbitrary asymptotically flat initial data) has
surjective linearization in these weighted Sobolev spaces. This result is proven
for maximal data by Choquet-Bruhat, Fischer, and Marsden in \cite{cfm}, and the
general case is due to Beig and \'{O} Murchadha \cite{bo:pg}.
\\

We now outline the basic approach to the main theorem.  Given asymptotically flat
initial data on $M$ satisfying (AC) in an appropriate chart at infinity in a given
end, we take a ``large'' radius $R$ and within the annulus from $R$ to $2R$, we
smoothly patch the given metric and second fundamental form to the metric and second
fundamental form coming from a slice in Kerr, or from another suitable family
satisfying (AC) near infinity.  This will produce an approximate solution of the
vacuum constraints.  The approximate solution is altered with a smooth perturbation
compactly supported within the closed annular gluing region to data
$(\bar{g},\bar{\pi})$, whose constraint function $\Phi(\bar{g},\bar{\pi})$ lies in
a finite-dimensional vector space of dimension ten; of course we want
$\Phi(\bar{g},\bar{\pi})$ to be zero.  (Note that outside the annulus the constraint
function vanishes by design.)  We then show that there is a choice of data to glue on so
that the resulting perturbed metric and second fundamental form are solutions to the
constraint equations.  The ten-dimensional obstruction space is countered by the ten
degrees of freedom afforded by
$(E,\mathbf{P},\mathbf{J},\mathbf{C})$; in fact the method proves that suitable models
near infinity are precisely those for which we can effectively attain values of
$(E,\mathbf{P},\mathbf{J},\mathbf{C})$ by moving within the family.  As in the previous
paper, the key to the proof is exploiting the overdetermined-ellipticity of the adjoint
of the linearization of $\Phi$; we include a fairly complete discussion of the analysis,
but some simple details are quoted from \cite{cor:schw}.  The conclusion of the proof is
the observation that the map from the parameter space to the obstruction space, a map
between two ten-dimensional Euclidean spaces, has non-zero degree; this map is more
complicated than before (where only four parameters were needed), but recognizing the
(AC) condition affords some economy in computing this map as compared to \cite{cor:schw},
where an expansion of the conformal factor is used. \\

We remark that the results of this paper are easily extended to allow $M$ to have
multiple asymptotically flat ends. For convenience of notation we write the proofs
under the assumption that $M$ has only one end.\\

The results of this paper (with the exception of Section 3) were announced in the
spring of 2000, and have been widely communicated to the mathematical GR community. In
the meantime, P.T. Chru\'{s}ciel and E. Delay \cite{cd} have obtained a version of these
results as well. Their paper employs our basic techniques (and those of the first
author in \cite{cor:schw}) to give a number of interesting applications; in addition
their paper gives an elegant and explicit description of the $10$-parameter Kerr family
of initial data which we do not do here. 

\section{Preliminaries.}
\subsection{Basic Notation.}

Let $\Omega$ denote a compactly contained domain in a smooth three-manifold $M^3$. 
Unless noted, we assume the boundary is smooth.  We list here some notation, and we
define some function spaces which we will find useful. 

\begin{enumerate}

\item[$\bullet$] Ric$(g)=R_{ij}$ and $R(g)=g^{ij}R_{ij}$ denote the Ricci and scalar
curvatures, respectively, of a Riemannian metric $g$ on $M$.  We use the Einstein
summation convention throughout, as well as the convention of using a semicolon to
denote covariant differentiation and a comma to denote partial differentiation, and our
convention for the Laplacian is $\Delta _{g}f
=\frac{1}{\sqrt{|g|}}\,\partial_{i}(g^{ij}\sqrt{|g|}\,\partial_{j}f)$.

\item[$\bullet$] We let $d\mu_g$ denote the volume measure induced by $g$,
$d\sigma_g$ the induced surface measure on submanifolds, $dx$ the Lebesgue
measure on Euclidean space, and $d\xi$ the Euclidean surface measure. 

\item[$\bullet$] We denote by $H^k$ the Hilbert space of tensor fields which
are square-integrable along with the first $k$ weak covariant derivatives, with the
standard $H^k$-inner product induced by the metric $g$.  It will be clear in context
which type of fields are being discussed. 
$H^k_{loc}$ denotes spaces of tensors which are in $H^k$ on each compact subset.  We
may abbreviate products with mixed regularity by using superscripts, like
$H^{2,1}=H^2\times H^1$.

Similarly we define the H\"{o}lder spaces $C^{k,\alpha}$.

\item[$\bullet$] We may wish to use symbols to distinguish various types of tensors. 
For example, let $\mathcal{M}^{k}$ ($k>\frac{3}{2}$) denote the open subset of $H^k$ of
Riemannian metrics, and $\mathcal{M}^{k,\alpha}$ denotes the open subset of metrics in
$C^{k,\alpha}$.  $\mathcal{S}_{(2,0)}$ denotes the space of symmetric (2,0)-tensor
fields, and $\mathcal{X}$ denotes the space of vector fields.  We use superscripts as
above to denote the desired Sobolev or H\"{o}lder regularity of the fields.

\item[$\bullet$] Let $\rho$ be a smooth positive function on $\Omega$.  Define
$L^{2}_{\rho}(\Omega)$ to be the set of locally $L^{2}(d\mu_g)$ functions $f$ such
that $f\rho^{1/2} \in L^{2}(\Omega,d\mu_g)$.  The pairing \[\langle
f_1,f_2\rangle_{L^2_{\rho}(\Omega)} = \langle
f_1\rho^{1/2},f_2\rho^{1/2}\rangle_{L^2(\Omega,d\mu_g)}\] makes $L^2_{\rho}(\Omega)$ a
Hilbert space.  It is clear how to extend this definition to higher order tensors.

\item[$\bullet$] Let $H^{k}_{\rho}(\Omega)$ be the Hilbert space of tensor fields in
$L^2_{\rho}(\Omega)$ along with the first $k$ covariant derivatives, the inner product
defined by incorporating the
$L^2_{\rho}(\Omega)$-pairings on the covariant derivatives of order $0,\ldots,k$.  

\item[$\bullet$] We now want to define the weighted H\"{o}lder spaces
$C^{k,\alpha}_{\rho^{-1}}(\Omega) \,\,\, (0<\alpha<1).$   We will consider $\rho$
which near $\partial \Omega$ decays as a power of or exponentially in the distance to
the boundary.  The weighted H\"{o}lder space is defined as the subspace of
$C^{k,\alpha}(\overline{\Omega})$ comprised of functions $f$ for which the norm
\[\|f\|_{k,\alpha,\rho^{-1}}:=\|f\rho^{-\frac{1}{2}}\|_{k,\alpha}\] is finite; this is
a Banach space.  (Unless noted
otherwise, norms will be taken over $\Omega$.)
\end{enumerate} 

The following lemma concerning the density of $H^k(\Omega)$ in $H^k_\rho(\Omega)$ will
be useful.
\begin{lemma}\label{density}
Assume that $\rho$ is bounded from above. For $k\geq 1$, the subspace $H^k(\Omega)$
(and hence $C^{\infty}(\overline{\Omega})$) is dense in $H^k_\rho(\Omega)$.
\end{lemma}
\noindent\emph{Proof:} For small $\tau>0$ let $\widetilde{\Omega}_\tau$ be a slightly
expanded domain compactly containing $\Omega$. Let $F_\tau:\widetilde{\Omega}_\tau\to
\Omega$ be a diffeomorphism which is $C^k$-close to identity map. Now suppose that
$f\in H^k_\rho(\Omega)$ and let $f_\tau=f\circ F_\tau$. Note that the restriction of
$f_\tau$ to $\Omega$ is in
$H^k(\Omega)$. We will show that $f_\tau\to f$ in $H^k_\rho(\Omega)$. First note that by
the chain rule we have $|D^\alpha f_\tau|\leq c\sum_{|\beta|\leq \alpha}|D^\beta f|\circ
F_\tau$ for $|\alpha|\leq k$. Now, the dominated
convergence theorem implies that $\lim\limits_{\delta\to 0^+}\int_{\Omega\setminus
\Omega_{\delta}}|D^\alpha f|^2\rho d\mu_g=0$ for $|\alpha|\leq
k$ where $\Omega_{\delta}$ denotes the subset of $\Omega$ consisting of points at
least a distance $\delta$ from the boundary. It follows that for $\tau>0$ we also have
$\lim\limits_{\delta\to 0^+}\int_{\Omega\setminus
\Omega_{\delta}}|D^\alpha f_\tau|^2\rho d\mu_g=0$. Thus, given $\epsilon>0$ we may
choose
$\delta$ so small that $(\int_{\Omega\setminus\Omega_{\delta}}|D^\alpha
f_\tau-D^\alpha f|^2\rho d\mu_g)^{1/2}<\epsilon/2$, and then choose $\tau$ so small that
$(\int_{\Omega_{\delta}}|D^\alpha f_\tau-D^\alpha f|^2\rho d\mu_g)^{1/2}<\epsilon/2$.
Summing these completes the proof. \endpf

\subsection{Linearization of the Constraint Map.}

We gather here a few facts we will use in what follows \cite{fm:def},
\cite{fm:defcon}.  

\begin{lemma} The scalar curvature map is a smooth Banach space map, as a map
$R:\mathcal{M}^{l+2}(\Omega) \rightarrow H^{l}(\Omega)$ ($l+2>\frac{5}{2}$), or
$R:\mathcal{M}^{k+2,\alpha}(\overline{\Omega})\rightarrow C^{k,\alpha}(\overline{\Omega})$ $(k\geq 0)$.  The
linearization $L_{g}$ of the scalar curvature operator is given by
\[L_{g}(h)=-\Delta_{g}(Tr_g(h))+div_g(div_g(h))-h\cdot Ric(g)\] in either of the
above spaces.  The formal $L^{2}$-adjoint $L_g^{*}$ of $L_g$ is given by 
\begin{eqnarray}
L_g^{*}(f)=-(\Delta _{g}f) g +Hess_g(f)-f\,Ric(g). 
\label{lemma:Radj}
\end{eqnarray}
\end{lemma}

\medskip

\begin{lemma}\label{conadj}  The constraint map $\Phi$ is smooth Banach space map, as a map
$\Phi:\mathcal{M}^{l+2}(\Omega)\times\mathcal{S}_{(2,0)}^{l+2}(\Omega) \rightarrow
H^{l}(\Omega)\times\mathcal{X}^{l+1}(\Omega)$ ($l+2>\frac{5}{2}$), or
$\Phi:\mathcal{M}^{k+2,\alpha}(\overline{\Omega})\times\mathcal{S}_{(2,0)}^{k+2,\alpha}(\overline{\Omega})\rightarrow
C^{k,\alpha}(\overline{\Omega})\times\mathcal{X}^{k+1,\alpha}(\overline{\Omega})$
($k\geq 0$).  The formal $L^{2}$-adjoint $D\Phi_{(g,\pi)}^*$ of the linearization
$D\Phi_{(g,\pi)}$ is given by $D\Phi_{(g,\pi)}^*(f,X)=
D\mathcal{H}_{(g,\pi)}^*(f)+Ddiv_{(g,\pi)}^*(X)$ where 
\begin{eqnarray*}
D\mathcal{H}_{(g,\pi)}^*(f)&=& \Big( (L_g^*f)_{ij} + \big( (Tr_g\pi)\pi_{ij}-2
\pi_{ik}\pi^{k}_{\, j} \big) f,\;\; \big( (Tr_g\pi) g^{ij}-2\pi^{ij}\big) f \Big) \\
Ddiv_{(g,\pi)}^*(X)&=& \frac{1}{2}\Big(  (L_X \pi)_{ij}+(X^k_{;k})\pi_{ij}-(X_i
\pi^k_{j;k} + \pi^k_{i;k} X_j ) \\ & & - X_{k;m}\pi^{km} g_{ij}-X_k\pi^{km}_{\,
;m}g_{ij},\;\; - (L_X g)^{ij} \Big).
\end{eqnarray*}


\end{lemma} 

\smallskip

\noindent We point out here that the key terms we will use are the $L_g^*f$-term in
the first component above, and the Lie derivative term $L_Xg$ in the second
component.   The formula is well-known and actually straightforward to derive, although
we note that in \cite{fm:defcon}, for example, the negative of the divergence operator
is used, so some signs differ; if we consider variations of $\pi$ as a (1,1)-tensor,
then the formula simplifies, as we will note in the next section. \endpf \\



We now indicate the source of the ten-dimensional obstruction space we mentioned in
the previous section.

\begin{lemma} Let $\Omega$ be an open domain in $\mathbb{R}^3$ with vacuum initial
data $(\delta,0)$.  Then the kernel $K$ of the operator $D\Phi_{(\delta,0)}^*$ is the
direct sum of the span $K_0$ of the functions $1$, $x^i$ ($i=1, 2, 3$) and the span
$K_1$ of the vector fields $X_i=\frac{\partial}{\partial x^i}$ ($i=1,2, 3$) and 
$Y_k=x^i\frac{\partial}{\partial x^j}-x^j\frac{\partial}{\partial x^i}$ ($i<j$, $k\neq
i,j$).
\label{lemma:coker}
\end{lemma}

\noindent\textit{Proof}:  By Lemma \ref{conadj}, $D\Phi_{(\delta,0)}^*(f,X)=\langle
L_{\delta}^*f, -\frac{1}{2}(L_X\delta)^{\sharp}\rangle $.  Moreover, by
Eq.(\ref{lemma:Radj}), $L_{\delta}^*f=0$ implies $Hess_{\delta}f=0$, $i.e.$ $f\in
K_0$.  That $L_X\delta =0$ means $X$ is a Killing field of the flat metric
on $\mathbb{R}^3$, so $X\in K_1$.  \endpf \\

We note that the kernel of $D\Phi^*_{(g,\pi)}$ has a natural interpretation.  We first recall that it is straightforward to show \cite{cor:schw} that a nontrivial element $f$ in the kernel of $L_g^*$
yields a warped product metric $-f^2 dt^2 +g$ that is Einstein, and if $R(g)=0$ then
this product metric is a solution to the Einstein vacuum equations which is manifestly
static.  Moncrief \cite{monc:stat} showed that a nontrivial
element $(f,X)$ in the the kernel of $D\Phi^*_{(g,\pi)}$ at a solution of the constraints corresponds to a Killing field (symmetry) in the resulting vacuum spacetime.


\section{Constructing Solutions with Good Asymptotics.}

In this section we show that any solution of the vacuum constraint equations with
$g_{ij}=\delta_{ij}+O(1/r)$ and $\pi^{ij}=O(1/r^2)$ may be perturbed by an arbitrarily
small amount on any compact set to a new solution satisfying the condition (AC). We
also show that the ADM energy-momentum vector is stable under this approximation in
that it is perturbed by an arbitrarily small amount. This construction shows that the
main theorem of the paper applies to a very large class of solutions of the vacuum
constraint equations which are dense in a suitable sense.\\

The construction we are going to make may be viewed as a generalization of the
deformation result of \cite{sy5} from the time-symmetric case to the case of arbitrary
data. We begin with any asymptotically flat data $(g,\pi)$ satisfying the vacuum
constraint equations and such that $g_{ij}=\delta_{ij}+O(1/r)$ and $\pi=O(1/r^2)$.
We are going to show how to approximate this data by new data $(\bar{g},\bar{\pi})$
satisfying the vacuum constraint equations and such that outside a compact
set we have
\begin{eqnarray}
\bar{g}_{ij}=u^4\delta_{ij},\ \bar{\pi}_{ij}=u^2({\cal L} X)_{ij}
\label{harmasym}
\end{eqnarray}
where $u$ tends to $1$ at infinity and $X$ is a vector field tending to $0$
at infinity, and where ${\cal L}$ is the operator related to the Lie derivative 
$L_X\delta$
$$ {\cal L}X=L_X\delta-div_\delta(X)\delta.
$$ 
If such asymptotics can be achieved, then the constraint equations near infinity
become the equations (computed with respect to $\delta$)
$$ 8 \Delta u= u\big( -|{\cal L}X|^2+1/2(Tr({\cal L}X))^2\big), \quad
 \Delta X^i+4u^{-1}u_j({\cal L}X)_{ji}-2u^{-1}u_i Tr({\cal L}X)=0 $$
where the second equation is written with respect to a euclidean basis. Standard
asymptotics (see \cite{ba:mass}) then imply that
$$ u(x)=1+a/r+O(1/r^2),\quad X^i=b^i/r+O(1/r^2)
$$
for constants $a,b^i$. These asymptotics clearly imply (AC).
\begin{thm} \label{harm}
Let $(g,\pi)$ be a vacuum initial data set as above. Given any $\epsilon>0$ and any
radius $R>>1$, there exists a vacuum initial data set $(\bar{g},\bar{\pi})$ satisfying
(\ref{harmasym}) which is within $\epsilon$ of $(g,\pi)$ on the ball of radius $R$.
Moreover, the mass and linear momentum of $(\bar{g},\bar{\pi})$ are within $\epsilon$
of those of $(g,\pi)$.
\end{thm}

In order to show that (\ref{harmasym}) can be achieved, we begin by modifying
$(g,\pi)$ to $(\hat{g},\hat{\pi})$ in the annular region from $R$ to $2R$ so that
$\hat{g}=\delta$ and $\hat{\pi}=0$ outside $B_{2R}$. We then attempt to reimpose
the constraint equations by constructing a solution of the form $\bar{g}=u^4\hat{g}$
and $\bar{\pi}=u^2(\hat{\pi}+{\cal L}X)$ where the operator ${\cal L}$ is computed
with respect the metric $\hat{g}$. We hope to find a solution $u$ which is near
$1$ and $X$ which is near $0$. The constraint equations then become the system
\begin{eqnarray}\nonumber
\bar{\mu}&=&u^{-5}[-8\Delta u+u(\hat{R}-|\hat{\pi}+{\cal
L}X|_{\hat{g}}^2+1/2(Tr_{\hat{g}}(\hat{\pi}+{\cal L}X))^2)]=0\\ \nonumber
(div_{\bar{g}}(\bar{\pi}))_i&=&u^{-2}[(div_{\hat{g}}(\hat{\pi}+{\cal
L}X))_i+4u^{-1}u_j(\hat{\pi}+{\cal L}X)_i^j-2u^{-1}u_iTr_{\hat{g}}(\hat{\pi}+{\cal
L}X)]=0.
\end{eqnarray}
We consider the map $T(u,X)=(\bar{\mu},div_{\bar{g}}(\bar{\pi}))$ and observe that
the linearization at $(1,0)$ is given by
$$ DT(\eta,Y)=(-8\Delta \eta-4\hat{\mu}\eta-4\hat{\pi}^{ij}Y_{i;j}+(div Y)
Tr\hat{\pi},\;\; div ({\cal L}Y)_i+4\eta_j\hat{\pi}^j_i-2\eta_iTr \hat{\pi}-2\eta
div_{\hat{g}}\hat{\pi})
$$  
where each term is computed with respect to $\hat{g}$. In order to solve the
equation
$T(u,X)=0$, we wish to use the inverse function theorem. Thus we would need to check
that $DT=DT_{(\hat{g},\hat{\pi})}$ is an isomorphism between appropriate spaces with
bounded inverse (independent of $R$ for large enough $R$). To be precise, we consider
$DT$ as an operator from
$W^{2,p}_{-\delta}$ to $W^{0,p}_{-\delta-2}$. The weighted norm convention we are
using is that the $W^{k,p}_{-\delta}$ norm is given by
$$ \|f\|_{k,p,-\delta}=\sum_{0\leq |\alpha|\leq k}(\int_M (|D^\alpha
f|\rho^{\delta+|\alpha|})^p\rho^{-3}d\mu_g)^{1/p}
$$
where $\rho$ is a function which equals $|x|$ near infinity and $\alpha$ is a
multi-index. It is well known that
$DT$ is a Fredholm operator of index $0$ for $p>1$ and $\delta\in (0,1)$ (see
\cite{ba:mass}). Thus in order to check that $DT$ is an isomorphism, it would suffice
to check that the cokernel is trivial; $i.e.$ that $DT$ is onto. This seems to be
difficult to check in general, so we use a less direct approach. We consider enlarging
the domain and we look for solutions of the modified problem
\begin{eqnarray}\nonumber
\Phi(u^4\hat{g}+h,u^2(\hat{\pi}+{\cal L}X)+k)=0
\end{eqnarray}
where $h$ and $k$ are suitably small symmetric $(0,2)$-tensors with {\it compact
support}. If we can solve this problem, the desired asymptotics follow as before. \\

We will need to use the fact that the operator $D\Phi=D\Phi_{(\hat{g},\hat{\pi})}$
is surjective as an operator from $W^{2,p}_{-\delta}\times W^{1,p}_{-\delta-1}$ to
$W^{0,p}_{-\delta-2}$  for $\delta\in (0,1)$. This result was proven by Choquet-Bruhat,
Fischer, and Marsden in \cite{cfm} in the maximal case, and by
Beig and \'{O} Murchadha \cite{bo:pg} in the general case. We give a direct proof in the next
proposition which appears to work under weaker asymptotic assumptions than those of
\cite{bo:pg}, although the same basic idea is employed.
\begin{prop} The operator $D\Phi_{(\hat{g},\hat{\pi})}$ is surjective from the domain
$W^{2,p}_{-\delta}\times W^{1,p}_{-\delta-1}$ onto $W^{0,p}_{-\delta-2}$ for $p>1$ and
$\delta\in (0,1)$.
\end{prop}  
\noindent\emph{Proof}: We will check the surjectivity for the initial data $(g,\pi)$,
and the result for $(\hat{g},\hat{\pi})$ will follow by a perturbation argument. We
first note the range of $D\Phi$ contains that of
$DT$ (a closed finite codimension subspace), and hence is of finite codimension in
$W^{0,p}_{-\delta-2}$. Therefore $D\Phi$ has closed range. If there is an element
$(\xi,Z)$ in the dual space $W^{0,q}_{-1+\delta}$ which annihilates the range of
$D\Phi$, then $(\xi,Z)$ lies in the kernel of the adjoint and hence satisfies the
equations: 
\begin{eqnarray}\nonumber
\xi_{;ij}-(\Delta \xi) g_{ij}-\xi
R_{ij}-1/2((K^{pq}Z_q)_{;p})g_{ij}+1/2(K_{ij}Z^p)_{;p}&=&0
\\ \nonumber -1/2(Z_{i;j}+Z_{j:i})+(div Z)g_{ij}-2\xi K_{ij}+2\xi (Tr K)g_{ij}&=&0. 
\end{eqnarray}
(Note that these are the adjoint equations with $K$ treated as a tensor of type $(1,1)$ so they differ slightly from those given in Lemma \ref{conadj}. The equations in this form can be found in \cite{cfm} for example, and are straightforward to derive.  For the application we have here, the detailed lower order terms do not play a role.)  We may take the trace and rewrite the equations eliminating the terms involving
$\Delta \xi$ and $div Z$. We then have

\begin{eqnarray}\nonumber
\xi_{;ij}-\xi R_{ij}+1/2(K_{ij}Z^p)_{;p}+[1/2R\xi+1/4(K^{pq}Z_q)_{;p}-1/4((Tr
K)Z^p)_{;p}]g_{ij}&=&0 \\ \nonumber 
1/2(Z_{i;j}+Z_{j:i})+2\xi K_{ij}&=&0.
\end{eqnarray}
Note also that taking the trace of the first equation and the divergence of the
second, we get a system of linear equations of the form $\Delta(\xi,Z)=B(x)(\nabla
\xi,\nabla Z)+C(x)(\xi,Z)=0$ where
$B,C$ are coefficient matrices. The following estimates are true for any 
weight $\tau>0$ and for any large radius $R$ 
\begin{eqnarray}\nonumber
\int_{M\setminus B_R}|\xi\rho^\tau|^2 \rho^{-3}d\mu_g&\leq&C\int_{M\setminus
B_R}(|\nabla\nabla\xi|\rho^{2+\tau})^2\rho^{-3}d\mu_g\\
\int_{M\setminus B_R}(|Z|\rho^\tau)^2\rho^{-3}d\mu_g&\leq& 
C\int_{M\setminus B_R}(|L_Zg|\rho^{1+\tau})^2\rho^{-3}d\mu_g
\label{locpoin}
\end{eqnarray}
where $(L_Zg)_{ij}=Z_{i;j}+Z_{j;i}$ is the Lie derivative. A proof of these
inequalities will be given below. Using these inequalities, we can
show inductively that
$(\xi,Z)$ must vanish to infinite order at infinity; $i.e.$  $|(\xi,Z)|\leq
C_N\rho^{-N}$ for any integer
$N>1$. To see this, we know from simple initial asymptotics for the equations that
$\xi$ and $Z$ are of order
$\rho^{-1}$. Putting this into the equations we have that $\nabla\nabla \xi$ is of
order $\rho^{-4}$ and $L_Zg$ is of order $\rho^{-3}$. Thus we may choose
$\tau<2$ and conclude that $(\xi,Z)$ are of order at most $\rho^{-\tau}$ for any $\tau<2$.
Note that the pointwise bounds follow from the corresponding weighted $L^2$ bounds by 
applying standard elliptic estimates (mean value-type inequalities) on balls of the 
form $B_{|x|/2}(x)$. 
Putting this information back into the equations, we find that $\nabla\nabla\xi$ is of
order
$\rho^{-\tau-3}$ and $L_Zg$ is of order $\rho^{-\tau-2}$. Thus we may choose $\tau<3$ and
improve the decay on $(\xi,Z)$. We thus conclude inductively that $(\xi,Z)$ vanishes to
infinite order at infinity.\\

To show that $(\xi,Z)$ vanishes identically, we use a standard unique continuation
result (see Kazdan \cite{k} for a version written in the most convenient form). To see
this, we consider doing an inversion (Kelvin transform). That is, we let $x=(x^1,x^2,x^3)$
denote asymptotically euclidean coordinates and we introduce $y=|x|^{-2}x$ where
$|x|$ is the euclidean norm of $x$. Now observe that the metric $\bar{g}=|x|^{-4}g$
has scalar curvature $\bar{R}=|x|^5(-8\Delta (|x|^{-1})+|x|^{-1}R)$ and this is of
order $|x|$ at infinity. We then observe that the metric $\bar{g}$ expressed in the
$y$ coordinates has Lipschitz components near $y=0$, and that
$\bar{g}_{ij}(0)=\delta_{ij}$. The conformal transformation for the Laplace operators
is then $\Delta u-1/8Ru=|y|^5(\bar{\Delta}(|y|^{-1}u)-1/8\bar{R}(|y|^{-1}u))$. It
follows that the quantities $(\bar{\xi},\bar{Z})=(|y|^{-1}\xi,|y|^{-1}Z)$ satisfy a
linear system of the form
$$\bar{\Delta}(\bar{\xi},\bar{Z})=
\bar{B}(y)(\bar{\nabla}\bar{\xi},\bar{\nabla}\bar{Z})+\bar{C}(y)(\bar{\xi},\bar{Z})
$$
with $\bar{B}$ is bounded near $y=0$ and $\bar{C}$ is bounded by a constant times
$|y|^{-1}$. The unique continuation theorem then implies that $(\xi,Z)$ vanishes
identically. \\

\noindent \emph{Proof of (\ref{locpoin})}: We choose a smooth cutoff function $\zeta$
which takes values between $0$ and $1$ with $\zeta(x)=0$ for $|x|\leq R$ and $\zeta(x)=1$
for $|x|\geq 2R$. The following inequalities for the euclidean metric $\delta$ on
$R^3$, and for functions $f$ and vector fields $X$ in the appropriate weighted spaces
are standard and can be proved using integration by parts
\begin{eqnarray}\nonumber
\int_M|f\rho^\tau|^2
\rho^{-3}dx&\leq&C\int_M(|\nabla f|\rho^{1+\tau})^2\rho^{-3}dx\\\nonumber
\int_M(|X|\rho^\tau)^2\rho^{-3}dx&\leq& 
C\int_M(|L_X\delta|\rho^{1+\tau})^2\rho^{-3}dx.
\end{eqnarray}
Thus we may take $f=\zeta \xi$ and $X=\zeta Z$ to obtain
\begin{eqnarray}\nonumber
\int_M|\zeta\xi\rho^\tau|^2
\rho^{-3}dx&\leq&C\int_M(|\nabla(\zeta\xi)|\rho^{1+\tau})^2\rho^{-3}dx\\\nonumber
\int_M(|\zeta Z|\rho^\tau)^2\rho^{-3}dx&\leq& 
C\int_M(|L_{\zeta Z}\delta|\rho^{1+\tau})^2\rho^{-3}dx
\end{eqnarray}
where all quantities are taken with respect to the euclidean metric on the end
$M\setminus B_R$. Using the choice of $\zeta$ and easy manipulations, these
inequalities clearly imply
\begin{eqnarray}\nonumber
\int_{M\setminus B_R}|\xi\rho^\tau|^2
\rho^{-3}dx&\leq&C\int_{M\setminus B_R}(|\nabla\xi|\rho^{1+\tau})^2\rho^{-3}dx+CR^{2\tau-3}\int_{B_{2R}\setminus B_R} \xi^2 dx\\ \nonumber
\int_{M\setminus B_R}(|Z|\rho^\tau)^2\rho^{-3}dx&\leq& 
C\int_{M\setminus
B_R}(|L_Z\delta|\rho^{1+\tau})^2\rho^{-3}dx+CR^{2\tau-3}\int_{B_{2R}\setminus
B_R}|Z|^2dx.
\end{eqnarray}
We may now prove (\ref{locpoin}) for the euclidean metric by contradiction. For
example, to prove the second inequality, suppose we have a sequence $Z_\alpha$ with 
$$ \int_{M\setminus B_R}(|Z_\alpha|\rho^\tau)^2\rho^{-3}d\mu_g=1
$$
but 
$$ \int_{M\setminus B_R}(|L_{Z_\alpha}g|\rho^{1+\tau})^2\rho^{-3}d\mu_g\to 0.
$$
We may assume that the $Z_\alpha$ converge weakly and in $L^2_{loc}$ to a limit $Z$. By
the inequality above we see that $Z\neq 0$, and we must have $L_Z\delta=0$. Therefore
$Z$ is a nonzero Killing vector field for $\delta$ with $\int_{M\setminus
B_R}(|Z|\rho^\tau)^2\rho^{-3}dv<\infty$. This is a contradiction which proves the second
inequality of (\ref{locpoin}) for the euclidean metric. By a similar argument one proves
$$ \int_{M\setminus B_R}|\xi\rho^\tau|^2 \rho^{-3}dx\leq C\int_{M\setminus
B_R}(|\nabla\xi|\rho^{1+\tau})^2\rho^{-3}dx,
$$
and the first inequality of (\ref{locpoin}) follows from applying this with $\xi$
replaced by $|\nabla\xi|$ and $\tau$ replaced by $\tau+1$. Using the fact that $g$
approaches the euclidean metric on approach to $\infty$, the inequalities
(\ref{locpoin}) follow. This completes the proof of the surjectivity of
$D\Phi$. \endpf \\

We are now in a position to complete the proof of Theorem \ref{harm}. It is well known
that $(h,k)\mapsto \Phi(g+h,\pi+k)$ is a continuously differentiable map from a neighborhood of
$(0,0)$ in
$W^{2,p}_{-\delta}\times W^{1,p}_{-1-\delta}$ to $W^{0,p}_{-2-\delta}$ for $p>3$ and
$\delta\in (0,1)$. We further observe that the truncated data $(\hat{g},\hat{\pi})$ are
arbitrarily close (for $R$ large) to $(g,\pi)$ in $W^{2,p}_{-\delta}\times
W^{1,p}_{-1-\delta}$. Thus the linearization $D\Phi$ is also surjective at
$(\hat{g},\hat{\pi})$. We may choose a basis $V_1,\ldots,V_N$ for the cokernel of
$DT$, and choose $(0,2)$-tensors $(h_1,k_1),\ldots, (h_N,k_N)$ in
$W^{2,p}_{-\delta}\times W^{1,p}_{-1-\delta}$  so that
$D\Phi(h_s,k_s)=V_s$ for $s=1,\ldots,N$. We now perturb the $(h_s,k_s)$ to
make them of compact support. The corresponding images under $D\Phi$ are close to the
$V_s$, and hence they still span a complementing subspace for the closed subspace
$Im (DT)$. We let $W_2$ be the linear span of $(h_1,k_1),\ldots,(h_N,k_N)$. We note that
the null space $U\subseteq W^{2,p}_{-\delta}$ of $DT$ is $N$-dimensional since the index of
$DT$ is $0$. We let $W_1$ be a closed complementing subspace to $U$, and we let
$W=W_1\times W_2$, so that $W$ is a Banach space. We then define the map $\overline{T}$ on
$W$ by setting $\overline{T}((u,X),(h,k))=\Phi(u^4\hat{g}+h,u^2(\hat{\pi}+{\cal L}X)+k)$. We
see by construction that $D\overline{T}$ is an isomorphism at $((1,0),(0,0))\in W$, so we
may apply the standard inverse function theorem to assert that $\overline{T}$ is an
isomorphism from a fixed (independent of $R$) neighborhood of $((1,0),(0,0))$ to a
fixed neighborhood of
$\Phi(\hat{g},\hat{\pi})$. It then follows that this image contains $(0,0)$, so we may find
$u$ near $1$, $X$ near $0$, $(h,k)$ near $0$ in the appropriate spaces so that
$\Phi(\bar{g},\bar{\pi})=0$ where $\bar{g}=u^4\hat{g}+h$ and
$\bar{\pi}=u^2(\hat{\pi}+{\cal L}X)+k$.\\

We now show that the ADM energy and linear momentum are perturbed by a small amount under
this perturbation of $(g,\pi)$. If $E,P_i$ denote the total energy and linear momentum of a solution $(g,\pi)$ of the vacuum constraint equations, we need only show that $E,P_i$ are
continuous with respect to the $W^{2,p}_{-\delta}\times W^{1,p}_{-1-\delta}$ norm for
$\delta>1/2$. For example, to see this for the $P_i$, observe that by the divergence
theorem we have for $R_1>R$ 
$$
\oint\limits_{|\mathbf{x}|=R_1}\pi_{ij}\nu^jd\sigma_g-
\oint\limits_{|\mathbf{x}|=R}\pi_{ij}\nu^jd\sigma_g=\int_{B_{R_1}\setminus B_R} \pi_{\
i;j}^jd\mu_g
$$
where $\pi_{\ i;j}^j$ represents the divergence of the {\it vector field} $\pi^j_{\
i}\partial/\partial x^j$. Using the constraint equation, we see that the volume integrand
is a sum of Christoffel symbol times components of $\pi$. Symbolically we write
$$ \int_{M\setminus B_R} |\Gamma||\pi|d\mu_g\leq \big(\int_{M\setminus
B_R}(|\Gamma|\rho^{1+\delta})^p\rho^{-3}d\mu_g\big)^{1/p}\big(\int_{M\setminus
B_R}(|\pi|\rho^{2-\delta})^q\rho^{-3}d\mu_g\big)^{1/q}.
$$ 
Now if $\delta>1/2$ we have
$$ \int_{M\setminus B_R}(|\pi|\rho^{2-\delta})^q\rho^{-3}d\mu_g\leq \big(\int_{M\setminus B_R}
(|\pi|\rho^{1+\delta})^p\rho^{-3}d\mu_g\big)^{q/p}\big(\int_{M\setminus
B_R}\rho^{-\epsilon-3}d\mu_g\big)^{(p-q)/p}
$$
where $\epsilon=(2\delta-1)pq/(p-q)>0$. Thus we have
$$ \Big| P_i-\oint\limits_{|\mathbf{x}|=R}\pi_{ij}\nu^jd\sigma_g\Big|\leq cR^{-\epsilon}
$$
where the constant $c$ depends only on the $W^{2,p}_{-\delta}\times W^{1,p}_{-1-\delta}$
norm of $(g-\delta,\pi)$. The continuity now follows from continuity of the
approximating surface integral. This completes the proof of Theorem \ref{harm}. \endpf

\section{Solving the Deformation Problem.}

In this section we prove the following local deformation theorem for the Einstein
constraints. 

\begin{thm}  Let $\Omega \subset M^3$ be a compactly contained $C^{k+2}$-domain, and let $g_{0}$ be a $C^{k+4,\alpha}$-metric
and $\pi_0$ a $C^{k+3,\alpha}$-symmetric (2,0)-tensor.  Suppose that the linearization
$D\Phi_{(g_0,\pi_0)}$ of the constraint map
$\Phi:C^{k+2,\alpha}(\Omega)\times\mathcal{S}_{(2,0)}^{k+2,\alpha}(\Omega)
\rightarrow C^{k,\alpha}(\Omega)\times \mathcal{X}^{k+1,\alpha}(\Omega)$ has an
injective formal $L^2$-adjoint  $D\Phi_{(g_0,\pi_0)}^{*}$ at $(g_{0},\pi_0)$, where 
we can consider \, $D\Phi_{(g_{0},\pi_0)}^{*}:H^{2,1}_{loc}(\Omega)
\rightarrow L^{2}_{loc}(\Omega)$.  Then for a sufficiently decaying weight $\rho$, there is an $\epsilon > 0$ such that for any
function $v\in C^{k,\alpha}(\overline{\Omega})$ and any vector field $W\in \mathcal{X}^{k+1,\alpha}(\overline{\Omega})$
for which $\big( (v,W)-\Phi(g_0,\pi_0)\big)\in C^{k,\alpha}_{\rho^{-1}}(\Omega)\times \mathcal{X}^{k+1,\alpha}_{\rho^{-1}}(\Omega)$ with
the support of $\big( (v,W)-\Phi(g_0,\pi_0)\big)$ contained in $\overline{\Omega}$ and
with $\|\big( (v,W)-\Phi(g_0,\pi_0)\big)\|_{C^{k,\alpha}_{\rho^{-1}}\times \mathcal{X}^{k+1,\alpha}_{\rho^{-1}}} < \epsilon$, there is
a $C^{k+2,\alpha}$-metric $g$ and a $C^{k+2,\alpha}$-symmetric
tensor $\pi$ on $M$ with $\Phi(g,\pi)=(v,W)$ in $\Omega$ and $(g,\pi)= (g_{0},\pi_0)$
outside $\Omega$.  Moreover, $(g,\pi)$ depends continuously on $(v,W)$.

If in addition $\big( (v,W)-\Phi(g_0,\pi_0)\big)\in C^{\infty}_c(\Omega)$, $(g_0,\pi_0)$ and $\partial \Omega$ are smooth, and we use an exponential weight, then we can solve for $(g,\pi)$ smooth.  

If the adjoint of the linearization has nontrivial kernel, then the analogous theorem holds for solving $\Phi(g,\pi)=(v,W)$ up to a finite-dimensional error.
\label{thm:locFM}
\end{thm}

The analogous result for the scalar curvature operator is found in \cite{cor:schw}; in
particular it is shown there that at generic metrics $g_0$, for functions $S$ so that the
difference $S-R(g_0)$ vanishes outside $\Omega$ and is sufficiently small (in a weighted H\"{o}lder norm), there is a metric $g$ with $R(g)=S$,
so that $g-g_0$ is small and supported in $\overline{\Omega}$.  The regularity statements are analogous to those above.\\

The proof of Theorem \ref{thm:locFM} will be carried out over the next few sections.  For simplicity we carry out the proof explicitly in the case we require to prove the main theorem, namely we have $(v,W)=0$ and we are solving up to a finite-dimensional error, as explained below; we will then remark on the straightforward modifications for the general case. 

\subsection{The Basic Estimate.}

We prove an elliptic estimate in the $\rho$-weighted Sobolev spaces in a $C^2$-domain
$\Omega$.  Since $\rho$ decays at $\partial\Omega$, the boundary decay is 
imposed by $\rho^{-1}$, not $\rho$, in the sense that tensors in $\rho^{-1}$-weighted
spaces will have to decay suitably at the boundary.  We remark that we get a global
estimate, even though the functions in the $\rho$-weighted spaces may not decay at the
boundary.  The key to achieving this estimate is using the overdetermined-ellipticity
correctly.  We state the estimate in the form needed for the asymptotic gluing we will do, and later remark how to get the estimate required for Theorem \ref{thm:locFM}.  \\

 We let $\zeta \in C^{\infty}_c(\Omega)$ be a bump
function, one on most of $\Omega$.  Let $K_*=\zeta K$ (recall that $K$ is defined in Lemma
\ref{lemma:coker}).  We define
$S_g$ to be the
$L^2(d\mu_g)$-orthogonal complement of $K_*$, and we take $S_0=S_{\delta}$.

\begin{thm}
There is a
constant $C$ and an $\epsilon >0$ so that for all data
$(g,\pi)$ within $\epsilon$ of the Minkowski data $(\delta,0)$ and for all $(f,X)\in
S_g$,
\begin{eqnarray}
\|(f,X)\|_{H^{2,1}_{\rho}(\Omega,d\mu_g)}\leq C
\|D\Phi_{(g,\pi)}^*(f,X)\|_{L^2_{\rho}(\Omega,d\mu_g)}.
\label{eq:BE} \label{thm:BE}
\end{eqnarray}
where the weight $\rho$ is $d^N$ or $e^{-1/d^N}$ near the boundary $\partial \Omega$
($N\geq 5$).
\end{thm}

\noindent\emph{Proof}:  We first prove estimates on the vector field $X$.  By Lemma
\ref{density} it suffices to prove these estimates for $X$ a smooth vector field on
$\overline{\Omega}$. Furthermore we define $\mathcal{L}_g X = L_X g$, to emphasize we
are thinking of the operator acting on $X$.
\begin{lemma}
For any weight function $\rho$ which is $d^N$ or $e^{-1/d^N}$ (with $N\geq 5$) near
the boundary $\partial \Omega$, there is a constant $C$ so that for all $X$ orthogonal 
(or merely in some fixed subspace transverse)
to the Killing fields of a metric $g$, the following estimate holds: 
\begin{eqnarray}
\|X\|_{H^1_{\rho}(\Omega)}\leq C \|\mathcal{L}_g X\|_{L^2_{\rho}(\Omega)}.
\end{eqnarray}
\end{lemma}

\noindent\emph{Proof}: We first note that since the Killing fields are Jacobi fields along
geodesics, they are locally determined by their first order jet at a point, and so the space
of Killing fields on $\Omega$ is finite-dimensional and the fields in the kernel will be
smooth (as much as the smoothness of the metric and domain will allow) up to the boundary
of $\Omega$.  The Lie derivative is an overdetermined-elliptic operator; that is, it has
injective symbol.  By standard elliptic theory, then, one gets an interior estimate of the
form
\[\|X\|_{H^1(\Omega_0)}\leq C (\|\mathcal{L}_g
X\|_{L^2(\Omega)}+\|X\|_{L^2(\Omega)})\] where $\Omega_0\subset\subset\Omega$ and $C$
depends on $d(\Omega_0,\partial\Omega)$.  To get a global estimate we foliate $\Omega$
near $\partial \Omega$ by level sets $\Sigma_r=\{ x\in\Omega :
d(x,\partial\Omega)=r\}$ of the distance function to $\partial \Omega$; there is an
$r_0>0$ depending on $\Omega$ for which these level sets are regular hypersurfaces. 
By standard elliptic theory on these closed hypersurfaces $\Sigma$ we get
\[\|X_{\Sigma}\|_{H^1(\Sigma)}\leq
C\left(\|\mathcal{L}_gX\|_{L^2(\Sigma)}+\|X\|_{L^2(\Sigma)}\right)\] since the
difference between $(\mathcal{L}_gX)|_{T\Sigma}$ and
$\mathcal{L}_{g_{\Sigma}}X_{\Sigma}$ is zero$^{th}$-order in $X$; $g_{\Sigma}$ is the
induced metric, and $X_{\Sigma}$ is the projection of $X$ onto $T\Sigma$.  The
constant $C$ in this estimate can be taken uniform in the distance to $\partial
\Omega$ sufficiently small.  Recall that since $|\nabla d(\cdot,\partial\Omega)|=1$,
the co-area formula gives for positive functions or $L^1$-functions $G$: $\int\limits_{r_1<d<r_2}G \; d\mu_g=\int\limits_{r_1}^{r_2}\int\limits_{\Sigma_r} G
\; d\sigma_g\;dr$.  Applying that here yields \begin{eqnarray*}
\|(X^T,\nabla_{\Sigma}X^T)\|_{L^2(A)}\leq
C\left(\|\mathcal{L}_gX\|_{L^2(A)}+\|X\|_{L^2(A)}\right) \end{eqnarray*} where $X^T$
is the vector field on $A=\{ x\in\Omega: r_1<d(x,\partial\Omega)<r_2\}$ given by the
projections $X_{\Sigma}$.  Since the constant is uniform in $r_1$, we see that
combining with the interior estimate we get
\[\|(X^T,\nabla_{\Sigma}X^T)\|_{L^2(\Omega_{\epsilon})}\leq
C\left(\|\mathcal{L}_gX\|_{L^2(\Omega_{\epsilon})}+\|X\|_{L^2(\Omega_{\epsilon})}\right)\]  
where $C$ is uniform in $\epsilon$ small, and $\Omega_{\epsilon}=\{x\in\Omega :
d(x,\partial\Omega)>\epsilon\}$.  We can use the co-area
formula to integrate the square of this estimate against $\rho^{\prime}(\epsilon)$ (which is
positive for $\epsilon$ small), and integrate by parts to get the desired weighted estimate, as in \cite{cor:schw}.\\

It remains to estimate the terms involving components of $X$ and $\nabla X$ normal to
$\Sigma$.  We compute in a local orthonormal frame $e_1,e_2,e_3$ adapted to $\Sigma$;
we take $e_3=\nabla d$, well-defined in a neighborhood of $\partial\Omega$.  We have
estimates for $X_{i;j}$ for $i,j=1,2$, and hence by taking the trace of
$\mathcal{L}_gX$, we can also estimate $X_{3;3}$ as above.  It suffices then to
estimate $X_{i;3}$ for $i=1,2$.   Let $E_i$ denote a set of the form $\{ x\in\Omega :
0<d(x,\partial\Omega)<r_i\}$, and let $\zeta$ be a cut-off function of the distance to
the boundary, which is identically one for $d<r_2<r_1\leq 1$, and vanishes outside $E_1$.  We first note  

\begin{eqnarray*}
2\int\limits_{E_2}\sum_{i=1}^2X_{i;3}^2 \rho d\mu_g & \leq &
2\int\limits_{\Omega}\zeta \sum_{i=1}^2X_{i;3}^2 \rho d\mu_g \\ &\leq &
\int\limits_{\Omega} |\mathcal{L}_gX|^2 \rho d\mu_g -2 \int\limits_{E_1} \zeta
\sum_{i=1}^2 X_{i;3}X_{3;i} \rho d\mu_g .\\ 
\end{eqnarray*}

We now want to integrate by parts to make the latter integrand $X_{i;3i}X_3$; we note
that by the arithmetic-geometric mean inequality (AM-GM) $ab\leq \frac{1}{2}\left(
\epsilon^2a^2+\frac{b^2}{\epsilon^2}\right)$, terms in the integrand of the form
$X_{i;3}X_j$ can be replaced (up to a constant factor) by $|X|^2$, with the other term being
absorbed on the left side of the above inequality.  This observation allows us to
integrate by parts and switch between covariant and partial derivatives at will. 
Since for $i=1,2$, $\zeta_{,i}=0=\rho_{,i}$, and since the boundary integrals vanish,
we get 
\begin{eqnarray*}
\int\limits_{E_2}\sum_{i=1}^2X_{i;3}^2 \rho d\mu_g & \leq & \int\limits_{\Omega}
|\mathcal{L}_gX|^2 \rho d\mu_g +2 \int\limits_{\Omega} \zeta \sum_{i=1}^2
X_{i;3i}X_{3} \rho d\mu_g + C \|X\|^2_{L^2_{\rho}(\Omega)}.\\ 
\end{eqnarray*}
We now use the Ricci formula to commute covariant derivatives: $X_{i;jk}-X_{i;kj}=X_lR^l_{kji}$. So we now have 
\begin{eqnarray*}
\int\limits_{E_2}\sum_{i=1}^2X_{i;3}^2 \rho d\mu_g & \leq & \int\limits_{\Omega}
|\mathcal{L}_gX|^2 \rho d\mu_g +2 \int\limits_{\Omega} \zeta \sum_{i=1}^2
X_{i;i3}X_{3} \rho d\mu_g + C \|X\|^2_{L^2_{\rho}(\Omega)}.
\end{eqnarray*}
If we integrate by parts again, we get 
\begin{eqnarray*}
\int\limits_{E_2}\sum_{i=1}^2X_{i;3}^2 \rho d\mu_g & \leq & \int\limits_{\Omega}
|\mathcal{L}_gX|^2 \rho d\mu_g -2 \int\limits_{\Omega} \zeta \sum_{i=1}^2
X_{i;i}X_{3;3} \rho d\mu_g \\ & & -2\int\limits_{\Omega} (\zeta\rho)_{,3} \sum_{i=1}^2
X_{i;i}X_{3} \rho d\mu_g + C\|X^T\|^2_{H^1_{\rho}(\Omega)} +C
\|X\|^2_{L^2_{\rho}(\Omega)}.
\end{eqnarray*}
We can estimate the second and third summands on the right side above by the AM-GM
inequality, bounding them by \[ C \int\limits_{\Omega}
|\mathcal{L}_gX|^2 \rho d\mu_g + C \|X\|^2_{L^2_{\rho}(\Omega)} + \int\limits_{\Omega}
|X|^2 d^{-2} \rho d\mu_g.\]

We now claim there is a constant $C$ (depending on $N$, as well as $g$ and $\Omega$)
so that \[\int\limits_{\Omega} |X|^2 d^{-2} \rho d\mu_g \leq C \int\limits_{\Omega}
|\mathcal{L}_g X|^2 \rho d\mu_g.\]  If this is not the case, we can find a sequence
$X_k$ of vector fields so that \[\int\limits_{\Omega} |X_k|^2 d^{-2} \rho d\mu_g =1\]
but for which \[\int\limits_{\Omega} |\mathcal{L}_gX_k|^2 \rho d\mu_g \rightarrow
0.\]  Now by the previous estimates, then, \[\int\limits_{\Omega} |\nabla_g X_k |^2
\rho d\mu_g \leq C \left( \int\limits_{\Omega} |\mathcal{L}_gX_k|^2 \rho d\mu_g +
\int\limits_{\Omega} |X_k|^2 \rho d\mu_g\right) +1.\]  Thus we see that
$X_k\rho^{1/2}$ is bounded in $H^1(\Omega)$, and so we can assume (by taking a
subsequence) that there is a vector field $X$ so that $X_k\rho^{1/2}$ converges weakly
in $H^1$ and strongly in $L^2$ to $X\rho^{1/2}$.  So $X$ is in $H^1_{loc}(\Omega)$,
and $X_k$ converges to $X$ locally in $L^2$, so that $\mathcal{L}_g X=0$ weakly.  But
the $X_k$ are transverse to the Killing fields, and so $X=0$, and thus 
$X_k\rho^{1/2}\rightarrow 0$ in $L^2$.  Now we note that there is an $r_1$ and a
constant $C$ so that on $E_1$, \[\Delta_g \rho \geq CN^2 d^{-2}\rho.\]  So we get by
integration by parts 
\begin{eqnarray}
N^2\int\limits_{\Omega} \zeta |X_k|^2 d^{-2}\rho d\mu_g &\leq & 2\int\limits_{\Omega}
\zeta |X_k||\nabla_gX_k|Nd^{-1}\rho d\mu_g + \int\limits_{\Omega}\zeta' |X_k|^2 d^{-1}
\rho.
\label{eq:est1}
\end{eqnarray}
Since $\zeta'$ is compactly supported in $\Omega$, the second integral on the right
side goes to zero in $k$.  By Cauchy-Schwarz we also have 
\begin{eqnarray*}
2\int\limits_{\Omega} \zeta |X_k||\nabla_gX_k|Nd^{-1}\rho d\mu_g & \leq & 
2N \left( \int\limits_{\Omega} \zeta |\nabla_gX_k|^2\rho d\mu_g \right)^{\frac{1}{2}}
\left( \int\limits_{\Omega} \zeta |X_k|^2 d^{-2}\rho d\mu_g\right)^{\frac{1}{2}}.
\end{eqnarray*}
Now using the previous estimate (\ref{eq:est1}), we get that 
\begin{eqnarray*}
N^2\int\limits_{\Omega} \zeta |X_k|^2 d^{-2}\rho d\mu_g &\leq & 2N (1+o(1))+o(1)
\end{eqnarray*}
where $o(1)$ denotes a function which goes to zero as $k\rightarrow \infty$.  So for $N$ at least 5, we
have for large $k$ \begin{eqnarray*}
\int\limits_{\Omega} \zeta |X_k|^2 d^{-2}\rho d\mu_g &\leq & \frac{1}{2}.
\end{eqnarray*}
But we also have that \[\int\limits_{\Omega}(1-\zeta)|X_k|^2d^{-2}\rho\] goes to zero
in $k$.  So far large enough $k$, \[ \int\limits_{\Omega}|X_k|^2 d^{-2}\rho d\mu_g <
1\] which is a contradiction. \endpf \\

We continue the proof of the basic estimate.  As we noted above (Lemma
\ref{lemma:coker}), the kernel of $D\Phi^*_{(\delta,0)}$ is $K_0\oplus K_1$.  $(f,X)\in S_0$ precisely if $f$ is orthogonal to $\zeta K_0$ and $X$ is
orthogonal to $\zeta K_1$.  By the preceding lemma, then, to prove the Basic Estimate at the
Minkowski data, we now just need to estimate $f$ by $L^*_{\delta}f$.  We do have an estimate of the form \[\|f\|_{H^2(\Omega)}\leq C \left(
\|L_g^*f\|_{L^2(\Omega)}+\|f\|_{L^2(\Omega)}\right).\] Indeed using
Eq.(\ref{lemma:Radj}) and taking a trace, we see that \[Hess_g f= -\frac{1}{2}\left(
Tr_g(L_g^*f)+f R(g) \right) g - f Ric (g).\]  Since $f$ is orthogonal to $\zeta K_0$, we can
apply standard theory (basically the Rellich theorem) to get the estimate 
\begin{eqnarray}
 \|f\|_{H^2(\Omega_{\epsilon})}&\leq & C \|L_g^*f\|_{L^2(\Omega_{\epsilon})}
\end{eqnarray}
where $C$ is uniform in $\epsilon$ small, and so we get the desired weighted estimate as above.\\

We now have the estimate for $(f,X)\in S_0$ 
\begin{equation}
\|(f,X)\|_{H^{2,1}_{\rho}(\Omega,\delta)}\leq
C\|D\Phi^*_{(\delta,0)}(f,X)\|_{L^2_{\rho}(\Omega,\delta)}.
\end{equation}
For data $(g,\pi)$ with $\epsilon$ of the Minkowski data, the estimate follows by perturbation.  \endpf


\subsection{Variational Method.}

In this section we show how solutions to
$\Pi_{S_0}D\Phi_{(g,\pi)}(h,\omega)=(\phi,V)$ for $(\phi,V)\in 
L^2_{\rho^{-1}}(\Omega) \cap S_0$, (and $(g,\pi)$ close to Minkowski data), can be obtained from
standard variational arguments.\\

We let $\Pi_{S_0}$ denote the orthogonal projection onto $S_0$ with respect to the
metric $\delta$, and we define the operator $\mathcal{R}_{(g,\pi)}$ to be the
linearization of the map $\Pi_{S_0}\circ \Phi$, so that
$\mathcal{R}_{(g,\pi)}= \Pi_{S_0} \circ D\Phi_{(g,\pi)}$.  We want to show this map
surjects from a suitable space onto $S_0\cap L^2_{\rho^{-1}}(\Omega)$.  In fact it is easy to see
this will be true if we can show the map $\mathcal{P}_{(g,\pi)}:=\Pi_{S_g} \circ
D\Phi_{(g,\pi)}$ is surjective to $S_g\cap L^2_{\rho^{-1}}(\Omega)$, for $(g,\pi)$
near the Minkowski data.\\  

We define the formal adjoint $\mathcal{P}^*_{(g,\pi)}$ with respect to the metric
$g$, and for $\rho(f,X)\in L^{2}_{\rho^{-1}}(\Omega)\cap S_g$ we define the functional $\mathcal{G}$ by 
\begin{eqnarray}
\mathcal{G}(u,Z)=\int\limits_{\Omega} \left( \frac{1}{2} \mid \mathcal{P}^*_{(g,\pi)}
(u,Z)\mid^2-(f,X)\cdot_g(u,Z)\right) \rho \; d\mu_g.
\end{eqnarray}   
We consider the infimum $\mu=\mu_{(f,X)}\leq 0$ over all $(u,Z)\in
\mathcal{V}_g=H^{2,1}_{\rho}(\Omega)\cap S_g$.  For any $(u,Z)\in \mathcal{V}_g$, and any
$(\psi,W)$ with compact support in $\Omega$, we have 
\begin{eqnarray}
\nonumber \langle \mathcal{P}^*_{(g,\pi)}(u,Z),(\psi,W)\rangle _{L^2(d\mu_g)}&=&
\langle (u,Z),\mathcal{P}_{(g,\pi)}(\psi,W)\rangle _{L^2(d\mu_g)}\\ \nonumber &=&
\langle (u,Z),D\Phi_{(g,\pi)}(\psi,W)\rangle _{L^2(d\mu_g)} \\ &=& \langle
D\Phi^*_{(g,\pi)}(u,Z),(\psi,W)\rangle _{L^2(d\mu_g)}
\label{eq:newadj}
\end{eqnarray}
where the second inequality follows from $(u,Z)\in \mathcal{V}_g$.  So on
$\mathcal{V}_g$ we have $D\Phi^*_{(g,\pi)}=\mathcal{P}^*_{(g,\pi)}$, and thus the
Basic Estimate will yield a minimizer to our variational problem, as we now recall.  We first note that $\mu$ is finite.

\begin{lemma}  For any $(f,X)\in L^2_{\rho}(\Omega),\, \mu>-\infty$. 
\end{lemma}
\textit{Proof}: We simply note that the Basic Estimate, along with Cauchy-Schwarz,
yields the estimate \[\mathcal{G}(u,Z)\geq
\frac{1}{2C}\|(u,Z)\|^2_{H^{2,1}_{\rho}(\Omega)}-\|(f,X)\|_{L^2_{\rho}(\Omega)}\|(u,Z)\|_{H^{2,1}_{\rho}(\Omega)}.\quad\quad
\endpf \]

\begin{cor} 
For any $(f,X)\in L^2_{\rho}(\Omega),\, \mu=\lim \limits_{i\rightarrow \infty}
\mathcal{G}(u_{i},Z_i)$ for some sequence $\{u_{i},Z_i\}$ with
$\{\|(u_{i},Z_i)\|_{H^{2,1}_{\rho}(\Omega)}\}$ bounded. \endpf
\end{cor}

Standard functional analysis now gives the existence of a minimizer $(u_0,Z_0)$ in the Hilbert space $\mathcal{V}_g$ \cite{cor:schw}.  In fact by the convexity of the functional
$\mathcal{G}$ the minimizer is unique.  In any case, since the orthogonal complement of $S_g$ is
composed of $C^{\infty}_c(\Omega)$-functions, we see that the Euler-Lagrange equations will
hold on all of $C^{\infty}_c(\Omega)$, yielding the weak formulation of 
\[ \mathcal{P}_{(g,\pi)}\rho\mathcal{P}^*_{(g,\pi)}(u_0,Z_0)=\Pi_{S_g}D\Phi_{(g,\pi)}\rho D\Phi^*_{(g,\pi)}(u_0,Z_0)=\rho(f,X).\]


A simple argument comparing the $L^2$-projections shows that $\rho
D\Phi^*_{(g,\pi)}(u_0,Z_0)$ is also weak solution of the equation 
\begin{eqnarray*} 
\Pi_{S_0}D\Phi_{(g,\pi)}\rho D\Phi^*_{(g,\pi)}(u_0,Z_0) = (\phi,V)
\end{eqnarray*}
where $(\phi,V)\in S_0 \cap L^2_{\rho^{-1}}(\Omega)$ and where we take 
$\mathcal{P}_{(g,\pi)}\rho D\Phi^*_{(g,\pi)}(u_0,Z_0)=\rho (f,X)$ to be the $L^2(d\mu_g)$-projection
of $(\phi,V)$ to $S_g$.  

\subsection{Pointwise estimates and the nonlinear problem.}

Assuming we start with tensors $(g_0,\pi_0)$ on $\mathbb{R}^3$, sufficiently smooth
and sufficiently close to Minkowski data on $\Omega\subset\subset\mathbb{R}^3$, and
which solve the constraint equations outside $\Omega$ (and in a neighborhood of the boundary), we produce tensors
$(\bar{g},\bar{\pi})$ which agree with the original data outside $\Omega$ and which
satisfy $\Pi_{S_0}\Phi(\bar{g},\bar{\pi})=0$.  The basic procedure is straightforward: we use the variational procedure to solve the
problem at the linear level, and we iterate the process of correction.  In this
section we write down the estimates needed to show convergence of the iteration.\\

We start by solving the equation $\Pi_{S_0}D\Phi_{(g_0,\pi_0)}\rho
D\Phi^*_{(g_0,\pi_0)}(u_0,Z_0) = -\Pi_{S_0}\Phi(g_0,\pi_0)$ as above.  Assuming
$(g_0,\pi_0) \in
\mathcal{M}^{k+4,\alpha}(\overline{\Omega})\times\mathcal{S}_{(2,0)}^{k+3,\alpha}(\overline{\Omega})$, we
get regularity on the solution $(u_0,Z_0)$ and interior elliptic estimates.  We could
explicitly derive these, but we will invoke the weighting method for elliptic systems
of mixed orders due to Douglis and Nirenberg \cite{dn}.  Note that for notational
simplicity we will omit the subscript $(g_0,\pi_0)$ on the operators below.\\

Now the operator $\rho^{-1}D\Phi\rho D\Phi^*$ is uniformly elliptic, and gives a
$4\times 4$ system, which we will symbolically write as $L_j U = L_{jk}(D)U^k$, and
the weights $s_j$ and $t_k$ are defined so that the order of $L_{jk}$ is 
$s_j+t_k$; the elliptic estimate comes from bounding the H\"{o}lder semi-norm $\sum\limits_{i=1}^4 \left[ U^i \right]_{t_i,\alpha}$ up to a constant factor by $\sum\limits_{i=1}^4 \left[ L_iU \right]_{-s_i,\alpha}$.  In our case $U$ stands for $(u,Z)$, and so we want the weights to satisfy the
following conditions: $s_1+t_1=4$; for $j,k>1$, $s_1+t_k=3=s_j+t_1$ and $s_j+t_k=2$. 
With this in mind we let $t_1=4$, and the other $t_k=3$, we let $s_1=0$ and the other
$s_j=-1$.  Then for $\Omega'\subset\subset\Omega$ 
\begin{eqnarray*}
\|(h_0,\omega_0)\|_{k+2,\alpha,\Omega'}&=& \|\rho
D\Phi^*(u_0,Z_0)\|_{k+2,\alpha,\Omega'} \leq  C \left(\|u_0\|_{k+4,\alpha,\Omega'} +
\|Z_0\|_{k+3,\alpha,\Omega'}\right)\\ &\leq & C\Big(
\|(u_0,Z_0)\|_{L^2_{\rho}(\Omega)}+\|D\mathcal{H}(\rho
D\Phi^*(u_0,Z_0))\|_{k,\alpha,\Omega}\\ & & + \|Ddiv(\rho
D\Phi^*(u_0,Z_0))\|_{k+1,\alpha, \Omega}\Big).    
\end{eqnarray*}  

If the weight $\rho$ is $d^N$ near the boundary, then the interior estimates local to
the boundary ($B'=B(x,\frac{d(x)}{4})\subset B=B(x,\frac{d(x)}{2})$) are
\begin{eqnarray*}
\|(h_0,\omega_0)\|_{k+2,\alpha,B'}&\leq& Cd^{N-\psi(k,\alpha)}\left(
\|(u_0,Z_0)\|_{L^2(B)}+\|\rho^{-1}D\mathcal{H}(\rho
D\Phi^*(u_0,Z_0))\|_{k,\alpha,B}\right.\\ & & + \left. \|\rho^{-1}Ddiv(\rho
D\Phi^*(u_0,Z_0))\|_{k+1,\alpha,B}\right).
\end{eqnarray*}
Here $\psi(k,\alpha)$ is linear in $k$ and $\alpha$, and also note that the lower-order coefficients in the operator $\rho^{-1} D\Phi \rho D\Phi^*$ have powers of $1/d$ that are accommodated by scaling in the estimate, or can also be subsumed into $\psi(k,\alpha)$.  Using the estimate \[\|(u_0,Z_0)\|_{L^2(B)}\leq
Cd^{-\frac{N}{2}}\|u_0,Z_0)\|_{L^2_{\rho}(\Omega)}\] we then have 
\begin{eqnarray*}
\|(h_0,\omega_0)\|_{k+2,\alpha,B'}&\leq&
Cd^{\frac{N}{2}-\psi(k,\alpha)}\Big(\|(u_0,Z_0)\|_{L^2_{\rho}}+\|\rho^{-\frac{1}{2}}D\mathcal{H}(\rho
D\Phi^*(u_0,Z_0))\|_{k,\alpha,B} \\ & &  +\|\rho^{-\frac{1}{2}}Ddiv(\rho
D\Phi^*(u_0,Z_0))\|_{k+1,\alpha,B}\Big).
\end{eqnarray*}

In fact we need to do the above estimates in terms of the projected operator. 
Because the difference between the operators $\rho^{-1}D\Phi\rho D\Phi^*$ and
$\rho^{-1}\mathcal{R}\rho D\Phi^*$ on $\mathcal{V}_g$ (see (\ref{eq:newadj})) is
finite-dimensional, we get the following interior estimate
\begin{eqnarray}
\|(h_0,\omega_0)\|_{k+2,\alpha,\Omega'}&=& \|\rho
D\Phi^*(u_0,Z_0)\|_{k+2,\alpha,\Omega'} \leq  C \left(\|u_0\|_{k+4,\alpha,\Omega'} +
\|Z_0\|_{k+3,\alpha,\Omega'}\right)  \nonumber \\ &\leq & C\left(
\|(u_0,Z_0)\|_{L^2_{\rho}(\Omega)}+\|\mathcal{R}\rho
D\Phi^*(u_0,Z_0)\|_{C^{k,\alpha}_{\rho^{-1}}(\Omega)\times\mathcal{X}^{k+1,\alpha}_{\rho^{-1}}(\Omega)}
\right).    
\label{eq:projint}
\end{eqnarray} 
Here $\Omega'$ is taken to be a ``large'' fixed domain containing the support of the
fields in $K_*$.  We have used that there is a constant $C$ so that for all
$(\phi,V)\in K_*$
\[\|(\phi,V)\|_{C^{k,\alpha}(\Omega)\times\mathcal{X}^{k+1,\alpha}(\Omega)}\leq C
\|(\phi,V)\|_{L^2(\Omega')}\] in particular for $(\phi,V)$ of the form $\Pi_{K_*}D\Phi\rho D\Phi^*(u,Z)$.  We then use that by compactness, for all $\epsilon>0$ there is a
$C(\epsilon)$ so that 
\[ \|(u,Z)\|_{C^{4}(\Omega')\times\mathcal{X}^{3}(\Omega')}\leq \epsilon
\|(u,Z)\|_{C^{k+4,\alpha}(\Omega')\times\mathcal{X}^{k+3,\alpha}(\Omega')}+C(\epsilon)\|(u,Z)\|_{L^2(\Omega')}.\] 
Since the fields in $\zeta K$ are supported away from the boundary,
$\mathcal{R}=D\Phi$ near the boundary.  Thus the estimates we have local to the
boundary are as in (\ref{eq:projint}), except for the appearance of the factor
$d^{\frac{N}{2}-\psi(k,\alpha)}$ on the right-hand side, which yields the decay of the
deformation tensors at $\partial\Omega$.  \\

Next we note that we can bound $(h_0,\omega_0)$ in $L^2_{\rho^{-1}}$ by the basic injectivity estimate for $D\Phi^*$, and we can bound the lower order term in the estimate by the variational inequality $\mathcal{G}(u_0,Z_0)\leq 0$, obtaining 
\[\|(h_0,\omega_0)\|_{L^2_{\rho^{-1}}} \leq C\|(u_0,Z_0)\|_{L^2_{\rho}}\leq C\|\Phi(g_0,\pi_0)\|_{L^2_{\rho^{-1}}}.\]
This inequality can be inserted into the decay estimates near the boundary, and it can also be used now to get a global estimate on $(h_0,\omega_0)$:
\[\|(h_0,\omega_0)\|_{L^2_{\rho^{-1}}}+\|(h_0,\omega_0)\|_{k+2,\alpha}\leq C \|\Phi(g_0,\pi_0)\|_{C^{k,\alpha}_{\rho^{-1}}\times\mathcal{X}^{k+1,\alpha}_{\rho^{-1}}}.\]   

We now iterate the process of linear correction.  We linearize
only about the initial metric and momentum tensor, since the coefficients in the
fourth-order system depend on derivatives of $(g_0,\pi_0)$; for example in computing
$L_g\rho L_g^*$, one differentiates the Ricci-term twice, which involves four derivatives of
the metric.  The above estimates show a gain of two derivatives in the metric deformation
tensor, not four.  Hence we cannot apply Newton's method to produce a solution.  However, we can still produce solutions by linear correction, as we now describe.  We use Taylor's formula in the form
\begin{eqnarray} 
\Phi(g_0+h,\pi_0+\omega)=\Phi(g_0,\pi_0)+D\Phi_{(g_0,\pi_0)}(h,\omega)+O(\|(h,\omega)\|_{k+2,\alpha}^2)
\end{eqnarray}
where
\begin{eqnarray}
\sup\limits_{(h,\omega)\neq 0}
\frac{\|O(\|(h,\omega)\|_{k+2,\alpha}^2)\|_{C^{k,\alpha}\times\mathcal{X}^{k+1,\alpha}}}{\|(h,\omega)\|^2_{k+2,\alpha}}=C(g_0,\pi_0)
\end{eqnarray}
and the constant can be taken uniformly for data near $(g_0,\pi_0)$ and on subdomains of $\Omega$.  We have the following Taylor's theorem for the projected operator on $\Omega$: 
\begin{eqnarray} 
\Pi_{S_0}\Phi(g_0+h,\pi_0+\omega)=\Pi_{S_0}\Phi(g_0,\pi_0)+\mathcal{R}_{(g_0,\pi_0)}(h,\omega)+O(\|(h,\omega)\|_{k+2,\alpha}^2).
\end{eqnarray}

By solving the equation \[\Pi_{S_0}D\Phi_{(g_0,\pi_0)}\rho
D\Phi^*_{(g_0,\pi_0)}(u_0,Z_0) = -\Pi_{S_0}\Phi(g_0,\pi_0)\] as above, then, we have
that \[\Pi_{S_0}\Phi(g_1,\pi_1)=O(\|(h_0,\omega_0)\|_{k+2,\alpha}^2)\] where
$(h_0,\omega_0)= \rho D\Phi^*_{(g_0,\pi_0)}(u_0,Z_0)$ and $(g_1,\pi_1)=
(g_0,\pi_0)+(h_0,\omega_0)$; this holds locally near the boundary as well, outside the support of $\zeta$.  The quadratic decay at the first step does not propagate
(since we linearize about the initial data only) but the estimates above show that a
Picard iteration converges; indeed it is straightforward to check that the error terms we get by using the fixed linear operator to correct the nonlinear term at each stage still allow the iteration to converge geometrically at a rate better than linear, but worse than quadratic, adapting the proof of Prop. 3.9 from \cite{cor:schw}.  (We remark that there are a few harmless typos in \cite{cor:schw}; in particular, $C^{k,\alpha}_{\rho^{-1}}$ should be defined as we have done here, and so the norm one uses on $(h,\omega)$ (simply $h$ in \cite{cor:schw}) in the Picard iteration is the norm on the left side of Eq.(\ref{eq:solbd}) below.)  The local estimates near the
boundary (outside the support of $\zeta$) allow us to show the limiting tensors can be made to decay as much as we like
as we approach $\partial\Omega$ by choosing $N$ large enough; we can even get all
derivatives decaying if we choose smooth data $\Phi(g_0,\pi_0)$ sufficiently small, supported away from
$\partial\Omega$, and if we use an exponential weight $\rho$.  Moreover we have a
bound on the limiting tensors $(h,\omega)=\sum\limits_{k=0}^{\infty}(h_k,\omega_k)$: 
\begin{eqnarray}
\|(h,\omega)\|_{L^2_{\rho^{-1}}}+\|(h,\omega)\|_{k+2,\alpha}\leq C \|\Phi(g_0,\pi_0)\|_{C^{k,\alpha}_{\rho^{-1}}\times\mathcal{X}^{k+1,\alpha}_{\rho^{-1}}}.
\label{eq:solbd}
\end{eqnarray}
This bound comes straight from the estimates done above (and the geometric convergence).\\

We remark that for the smooth case we solve the nonlinear equation in some finite regularity class to start, but using an exponentially decaying weight.  We use standard bootstrapping on the quasilinear elliptic condition $\Phi((g_0,\pi_0)+\rho D\Phi^*(u,Z))\in K_* \subset C^{\infty}$ to see that the solution $\rho D\Phi^*(u,Z)$ is smooth on the interior, and then use the decay of $\rho$ to prove that all the derivatives decay near the boundary; we note that we have assumed that $\Phi(g_0,\pi_0)$ is supported away from the boundary, and we are considering the case $(v,W)=0$ of Theorem \ref{thm:locFM}.  The general case is similar, as we discuss in the next remark. \\
 
\noindent \textit{Remark}.  We now note what modifications are needed to carry out the above argument to prove Theorem \ref{thm:locFM} as stated above.  Essentially we just need to derive the analogous Basic Estimate.  The variational method and nonlinear iteration will proceed as above.  In fact, when there is no cokernel present, the method is slightly easier, since we do not need to worry about staying transverse to a subspace.  The Basic Estimate follows by first observing the estimate $\|X\|_{H^1_{\rho}(\Omega)}\leq C \big(\|\mathcal{L}_g X\|_{L^2_{\rho}(\Omega)} + \|X\|_{L^2_{\rho}(\Omega)}\big)$ follows by combining the estimate above with the fact mentioned earlier that the kernel of $\mathcal{L}_g$ is finite-dimensional and smooth up to the boundary.  Note that by differentiating the momentum constraint (and using the Ricci formula) we see that the condition that $(f,X)$ is in the kernel of $D\Phi^*$ reduces to a second-order ODE system along geodesics, and so the kernel is also finite-dimensional and smooth (as smooth as allowed by the other data).  Elementary manipulation using Lemma \ref{conadj} then yields the estimate 
\[ \|(f,X)\|_{H^{2,1}_{\rho}(\Omega)}\leq
C\Big( \|D\Phi^*_{(g,\pi)}(f,X)\|_{L^2_{\rho}(\Omega)}+ \|(f,X)\|_{L^2_{\rho}(\Omega)}\Big) .\]  
We can remove the lower-order term in the estimate by essentially the usual Rellich argument; to do this, one first shows the left side of the above estimate can be replaced by $\|(f,X)\rho^{\frac{1}{2}}\|_{H^2(\Omega)}$.  In fact we already have the required estimate for the vector field $X$ above, and the analogous estimate for $f$ in terms of $L^*_gf$ is a straightforward modification; if we had incorporated different weights on the derivatives in the definition of the weighted spaces, this would have already been included in the Basic Estimate.   This completes the proof of Theorem \ref{thm:locFM}.  \endpf

\section{Handling the cokernel.}

We apply the preceding analysis to solve the constraint equations with given model near infinity 
up to a finite-dimensional obstruction.  Given any AF solution of the constraints on
$\mathbb{R}^3$, we fix AF coordinates at infinity, and at a sufficiently large radius
$R$ we smoothly patch (using a smooth cut-off function) our original data to the model 
data, the slow-patching occurring in the annulus $A_R$ from $R$ to $2R$.  To apply the
preceding section, we scale our glued data to the unit annulus $A_1$ by $(g_R(x),\pi_R(x)):=(g(Rx),R\pi(Rx))$.  Under this scaling the relevant geometric
quantities (derivatives, curvatures, etc.) are $O(R^{-1})$, the (AC) conditions hold to order $O(R^{-2})$, and the vacuum
constraint equations are scale-invariant.  For large enough $R$ the data will be sufficiently close to the flat
data on the fixed annulus $A_1$, so the data can be perturbed to make the constraint functions lie in $\zeta K$ on $A_1$.\\

It is at this point we need to be clear about what are suitable families of solutions to glue on near infinity.  We define a family of solutions on the exterior of a fixed ball and smoothly parametrized on an open set $\mathcal{O}\subset \mathbb{R}^{10}$ to be \textit{admissible} if, with reference to a fixed coordinate chart near infinity, the family satisfies (AC) locally uniformly, and the map $\Theta:\mathcal{O}\rightarrow \mathbb{R}^{10}$ which associates to each member of the family its energy-momenta $(E,\mathbf{P},\mathbf{J}.\mathbf{C})$ is a homeomorphism onto an open subset.\\

We note that slices in Kerr form an admissible family,
parametrized by the total mass $m$, the angular momentum parameter $a$, and an element in the Poincar\'{e} group to control Euclidean motions of the AF coordinate system as well as boosts.  It is well-known that under the action of the Poincar\'{e} group the energy-momentum transforms as a four-vector, and the center and angular momentum enjoy a similar transformation rule (they comprise a skew-symmetric $M_{\mu\nu}$ with $M_{0k}$ a constant times $C^k$) $cf.$ \cite{adm:mass}, \cite{bo:pg}, \cite{rt:RT}, which allows one to see that varying the slices near the given one gives a local homeomorphism from a ten-dimensional family of slices to the ten-dimensional space of energy-momenta.  Note that time translations and rotations about the axis of symmetry of the Kerr are isometries, so we really mod out by the closed two-parameter subgroup they generate to parametrize the slices, leaving eight effective parameters from the group.  Please see \cite{cd} for an explicit formula and other examples. \\

Let $E_R\subset M$ correspond the the exterior $\{ x\in \mathbb{R}^3: |x|>R \}$ in an AF chart.  We now state the Main Theorem:

\begin{thm}\label{main} Let $(g,\pi)$ solve the vacuum constraint equations on $M^3$, and satisfy $(AC)$, in an AF coordinate chart in a given end.  Let $\mathcal{O}$ parametrize an admissible family of solutions, and $\lambda_0\in\mathcal{O}$ has $\Theta(\lambda_0)$ matching that of $(g,\pi)$.  There is a radius $R$ and a solution $(\overline{g},\overline{\pi})$ of the constraints so that $(\overline{g},\overline{\pi})=(g,\pi)$ on $M\setminus E_R$, and $(\overline{g},\overline{\pi})$ agrees with a suitably chosen member of the admissible family on $E_{2R}$.\end{thm}

\noindent Coupled with Theorem \ref{harm}, we have the following approximation result.

\begin{thm}\label{main2} Let $(g,\pi)$ be any AF solution of the vacuum constraints.  Given any $\epsilon >0$, there is a solution $(\overline{g},\overline{\pi})$ within $\epsilon$ of $(g,\pi)$ and whose ADM energy-momentum $(E,\mathbf{P})$ is within $\epsilon$ of that of $(g,\pi)$, so that near infinity, $(\overline{g}, \overline{\pi})$ agrees with a member of an admissible asymptotic model family.
\end{thm}

\noindent \textit{Proof of Theorem \ref{main}}: For sufficiently large $R$, we have a continuous map $\mathcal{I}$ from an open set $\mathcal{O} \subset \mathbb{R}^{10}$ to $\mathbb{R}^{10}$ as follows:
take data corresponding to $\lambda\in\mathcal{O}$, glue it to the given data in $A_R$, and
then perturb using the above techniques to data $(\bar{g},\bar{\pi})$ with $\Phi(\bar{g},\bar{\pi})$ lying in a fixed ten-dimensional vector space; we identify $\Phi(\bar{g},\bar{\pi})$ with $\mathcal{I}(\lambda)$.  By design, $(\bar{g},\bar{\pi})$ is identical to the member of the asymptotic model family in the exterior end $E_{2R}$.  In the next section we show that this map $\mathcal{I}$ has a zero near the parameter $\lambda_0 \in\mathcal{O}$ corresponding to the given initial AF
data $(g,\pi)$ with which we started, and thus this data solves the vacuum constraints.  Notice that we cannot arbitrarily glue on anything we like, but the procedure finds a suitable exterior that will lead to a solution of the constraints.
  
\subsection{Computing the parameter map.} 

In this section we verify that the compactly supported deformation which puts the
constraint data into the approximate kernel preserves the asymptotic symmetric
conditions (AC).  This allows us to conclude that the projection of the data onto the
cokernel is given to leading order by the change in parameters across the annulus from
the initial to the model we glued on; $i.e.$ the deformation does not induce any extra
error terms.  This simplifies the computations in \cite{cor:schw}, where less precise
estimates were used along with explicit calculations in the conformally flat case. 

\subsubsection{Preservation of the AC Condition}

Starting with $g$ and $\pi$ satisfying the asymptotic condition
(AC) (in an AF chart, including several derivatives), we glue on a model solution on $A_R$, using a symmetric cut-off function which is zero near one of the boundary spheres, and one near the other; this produces an approximate solution $(\widetilde{g},\widetilde{\pi})$.  We note that the model solutions which we consider satisfy (AC), uniformly near a given one.  Next, we scale to the unit annulus
$A_1$: $(g_R)_{ij}(x):=\widetilde{g}_{ij}(Rx)$, and $(\pi_R)_{ij}(x):=R\widetilde{\pi}_{ij}(Rx)$.  We
see that $(g_R)_{ij}(x)-(g_R)_{ij}(-x)=O(R^{-2})$,
$(\pi_R)_{ij}(x)+(\pi_R)_{ij}(-x)=O(R^{-2})$, and similarly the derivatives will
satisfy the appropriate even/odd condition to order $O(R^{-2})$, and hence the
Christoffel symbols and curvatures do as well.  We use the notation
$\widehat{u}(x)=u^{\wedge}(x)=u(-x)$, and we also note that ``$O$" will include several derivatives of the quantity in question (as required in the next section), and on
the annulus $A_1$ the derivatives will decay at the same rate by the scaling.\\

Let $(h,\omega)=\rho D\Phi^*_{(g_R,\pi_R)}(u,Z)$ solve the nonlinear projected
problem; in particular, $(h,\omega)$ is obtained by iteration, so that
$(h,\omega)=\sum\limits_{k=0}^{\infty} (h_k,\omega_k)$, where
$(u,Z)=\sum\limits_{k=0}^{\infty} (u_k,Z_k)$ and $(h_k,\omega_k)=\rho
D\Phi^*_{(g_R,\pi_R)} (u_k,Z_k)$, and \[\Phi(g_R+h,\pi_R + \omega)=\Big(
\sum\limits_{i=0}^3c_i x^i \zeta, \sum\limits_{j=1}^3 a_j X_j \zeta +
\sum\limits_{k=1}^3 b_k Y_k \zeta\Big) \] on $A_1$, with $x^0:=1$.  Since
$\Phi(g_R,\pi_R)=O(R^{-1})$, we have by (\ref{eq:solbd}) that $(h,\omega)=O(R^{-1})$.\\

We now note that the anti-symmetric part of $\Phi(g_R+h,\pi_R+\omega)$ is small. 
Indeed, for $i=1,2,3$ we have $c_i=O(R^{-2})$, $b_i=O(R^{-2})$:
\begin{eqnarray*}
\int_{A_1}\mathcal{H}(g_R+h,\pi_R+\omega)x^i \; dx &=& \int_{A_1} \big[
R(g_R)+L_{g_R}(h)+O(\|h\|_{2,\alpha}^2)+ O(R^{-2}) \big] x^i \; dx \\ &=& \int_{A_1^+}
\big[R(g_R)-\widehat{R(g_R)} \big] x^i \; dx + \int_{A_1} L_{g_R}(h)x^i d\mu_{g_R} +
O(R^{-2})
\end{eqnarray*}
where we have used the fact that $|dx-d\mu_{g_R}| = O(R^{-1})$.  The first term above
is $O(R^{-2})$ by the symmetry of $R(g_R)$, as is the second integral, since
$L_{g_R}^*(x^i)=O(R^{-1})$; indeed we note \[\big(Hess_{g_R}x^i \big)_{jk} = -
\Gamma^m_{jk}(dx^i)_m=O(R^{-1}).  \]   Similarly, using the symmetry of $Y_k$
\begin{eqnarray*}
\int_{A_1}div_{g_R+h} (\pi_R+\omega)\cdot Y_i \; dx &=& \int_{A_1}
div_{\delta}(\pi_R+\omega) \cdot Y_i \; dx +O(R^{-3})= O(R^{-2}).
\end{eqnarray*}

It directly follows that
$D\Phi_{(g_R,\pi_R)}(h,\omega)(x)-D\Phi_{(g_R,\pi_R)}(h,\omega)(-x)=O(R^{-2})$.  With
a little computation, we also see that
$D\Phi_{(g_R,\pi_R)}(h-\widehat{h},\omega+\widehat{\omega})$,  $\rho\Big(
\big(D\Phi_{(g_R,\pi_R)}^*(u,Z)\big)^{\wedge}-
D\Phi_{(g_R,\pi_R)}^*\widehat{(u,Z)}\Big)$, and hence $D\Phi_{(g_R,\pi_R)}\rho
D\Phi_{(g_R,\pi_R)}^*(u-\widehat{u},Z-\widehat{Z})$, are also $O(R^{-2})$.   Here we use the fact that
we have uniform bounds on $\rho \nabla^k u$ ($k=0,1,2$) and $\rho \nabla ^k Z$
($k=0,1$) and their derivatives from the elliptic estimates (and the decay of $\rho$).\\

Now we estimate $(h-\widehat{h},\omega+\widehat{\omega})= \rho\Big(
D\Phi_{(g_R,\pi_R)}^*(u,Z)-\big( D\Phi_{(g_R,\pi_R)}^*(u,Z)\big)^{\wedge} \Big) +
O(R^{-2})=\rho\Big(
D\Phi_{(g_R,\pi_R)}^*(u,Z)-D\Phi_{(g_R,\pi_R)}^*\widehat{(u,Z)}\Big) +O(R^{-2})$.   
By the elliptic estimate, using the interior estimate near the boundary to
establish decay and global bounds, we see that it suffices to show that
$(u,Z)-\widehat{(u,Z)}$ is $O(R^{-2})$ in $L^2_{\rho}(A_1)$.  To show this, we note
that the above symmetry discussion, coupled with linearization and the decay estimates
obtained from the elliptic estimate, shows that $D\Phi_{(g_R,\pi_R)}\rho
D\Phi_{(g_R,\pi_R)}^*(u-\widehat{u},Z-\widehat{Z}) \in L^2_{\rho^{-1}}(A_1)$ and that
we can write  \[\int_{A_1}\big( (u,Z)-\widehat{(u,Z)}\big) D\Phi_{(g_R,\pi_R)}\rho
D\Phi_{(g_R,\pi_R)}^*(u-\widehat{u},Z-\widehat{Z}) d\mu_{g_R} =
\|(u,Z)-\widehat{(u,Z)}\|_{L^2_{\rho}} O(R^{-2}).\]  Furthermore, $(u,Z)$ and hence
$\widehat{(u,Z)}$ are transverse to $K_*$.  This allows us to apply 
integration by parts (using Lemma \ref{density}) and the basic elliptic estimate transverse to the cokernel to get
\begin{eqnarray*}
 \int_{A_1}\big( (u,Z)-\widehat{(u,Z)}\big) D\Phi_{(g_R,\pi_R)}\rho
D\Phi_{(g_R,\pi_R)}^*(u-\widehat{u},Z-\widehat{Z}) d\mu_{g_R} & &\\ = \int_{A_1} \rho
|D\Phi_{(g_R,\pi_R)}^*(u-\widehat{u},Z-\widehat{Z})|^2 d\mu_{g_R} &\geq& C
\|(u,Z)-\widehat{(u,Z)}\|^2_{H^{2,1}_{\rho}(A_1)}.
\end{eqnarray*}

\subsubsection{Controlling the resulting constraint data.}

The preservation of the (AC) condition implies that the projection of the constraint
functions in the direction of the cokernel after solving the nonlinear projected
problem is given, up to lower-order error terms, by boundary integrals; of course the
boundary data \textit{and its derivatives} are unchanged by the perturbation in the
annulus.  In particular, this means that the constraint functions can be controlled by
varying the parameters of the solution we glue on, and so as long as the family we
glue on has enough degrees of freedom, we can make the Hamiltonian and momentum
constraints zero.  Indeed, we let $(\overline{g}_R,\overline{\pi}_R)=(g_R+h,\pi_R +
\omega)$, we let $(\overline{g},\overline{\pi})$ be the re-scaled version on $A_R$,
and we note for $k=1,2,3$
\begin{eqnarray*} 
\int_{A_1} x^k \mathcal{H}(\overline{g}_R,\overline{\pi}_R) dx &=& \int_{A_1} x^k
\sum\limits_{i,j} \big[ (\overline{g}_R)_{ij,ij}-(\overline{g}_R)_{ii,jj} \big] dx +
O(R^{-3}) \\ &=& R^{-2} \int_{A_R} x^k \sum\limits_{i,j}\big(
\overline{g}_{ij,ij}-\overline{g}_{ii,jj} \big) dx + O(R^{-3})\\ &=& R^{-2}
\int_{\partial A_R} x^k\sum\limits_{i,j} \big( \tilde{g}_{ij,i}-\tilde{g}_{ii,j} \big)
\nu^j d\xi \\ & & - R^{-2} \int_{\partial A_R} \sum\limits_{i}\big(\tilde{g}_{ik} \nu^i
- \tilde{g}_{ii}\nu^k \big)  d\xi + O(R^{-3}).
\end{eqnarray*}
Here we have used the fact that the difference between $R(\overline{g}_R)$ and
$\sum\limits_{i,j}\big[ (\overline{g}_R)_{ij,ij}-(\overline{g}_R)_{ii,jj}\big]$ is
$O(R^{-2})$ and is even to $O(R^{-3})$, and similarly for the quadratic terms in
$\overline{\pi}_R$ in the Hamiltonian constraint.  We now note the other projections,
starting with the mass term, the projection onto the constant functions:
\begin{eqnarray*}
\int_{A_1} \mathcal{H}(\overline{g}_R,\overline{\pi}_R) dx &=& \int_{A_1}
\sum\limits_{i,j} \big[ (\overline{g}_R)_{ij,ij}-(\overline{g}_R)_{ii,jj} \big] dx +
O(R^{-2}) \\ &=& R^{-1} \int_{A_R} \sum\limits_{i,j}\big(
\overline{g}_{ij,ij}-\overline{g}_{ii,jj} \big) dx + O(R^{-2})\\ &=& R^{-1}
\int_{\partial A_R} \sum\limits_{i,j} \big( \tilde{g}_{ij,i}-\tilde{g}_{ii,j} \big)
\nu^j d\xi + O(R^{-2}).
\end{eqnarray*}
Integrating the momentum constraint against the translation fields $X_k$ yields 
\begin{eqnarray*}
\int_{A_1} div_{\overline{g}_R} (\overline{\pi}_R)\cdot X_k \; dx &=& \int_{A_1}
div_{\delta}(\overline{\pi}_R) \cdot X_k \; dx +O(R^{-2})\\ &=& \int_{\partial A_1}
(\overline{\pi}_R)_{ij} \cdot X^i_k \nu^j \; d\xi + O(R^{-2})\\ &=&  R^{-1}
\int_{\partial A_R} \overline{\pi}_{ij}X_k^i \nu^j\; d\xi + O(R^{-2}).
\end{eqnarray*}
Finally, we project the momentum constraint onto the rotation fields $Y_k$:
\begin{eqnarray*}
\int_{A_1} div_{\overline{g}_R} (\overline{\pi}_R)\cdot Y_k \; dx &=& \int_{A_1}
div_{\delta}(\overline{\pi}_R) \cdot Y_k \; dx +O(R^{-3})\\ &=& \int_{\partial A_1}
(\overline{\pi}_R)_{ij} \cdot Y^i_k \nu^j \; d\xi + O(R^{-3})\\ &=&  R^{-2}
\int_{\partial A_R} \overline{\pi}_{ij}Y_k^i \nu^j\; d\xi + O(R^{-3}).
\end{eqnarray*}

Note that we could also have used the measures from the metric $\overline{g}_R$ in these integrals.  Also recall that $(\overline{g},\overline{\pi})$ agree with the given data $(g,\pi)$ on the inner boundary, and agree with the data of the asymptotic model on the outer boundary of the
annulus.  To show the map $\mathcal{I}$
has a zero, we show the associated map $\mathcal{I}_R: \lambda \mapsto$ 
\[\Big( R\int_{A_1} \mathcal{H}(\overline{g}_R,\overline{\pi}_R) dx , R\int_{A_1}
div_{\overline{g}_R} (\overline{\pi}_R)\cdot X_k , R^2\int_{A_1} div_{\overline{g}_R}
(\overline{\pi}_R)\cdot Y_k \; dx, R^2\int_{A_1} x^k
\mathcal{H}(\overline{g}_R,\overline{\pi}_R) dx \Big)\] has a zero.  

\subsection{Solving the constraints.}

We now finish the proof of the Main Theorem.  Thus far, we have not chosen which member of the
model family ($e.g.$ which slice in which Kerr) to use near infinity.  We understand that
the model will have ADM energy-momentum, and angular and linear momentum, near that of the
original data.  Suppose $\lambda_0 \in\mathcal{O}$ parametrizes a solution whose
energy-momenta $\Theta(\lambda_0)$ agrees with that of $(g,\pi)$.  We then consider
$\lambda$ near $\lambda_0$.\\

The computation in the previous section shows then that the map
$\mathcal{I}_R(\lambda)=\Theta(\lambda)-\Theta(\lambda_0) +o(1)$, where $o(1)\rightarrow 0$
as $R\rightarrow \infty$, uniformly for $\lambda$ near $\lambda_0$.  Fix a small ball $B$
about $\lambda_0$.  Let $\mathcal{I}_R(t,\lambda)= \Theta(\lambda)-\Theta(\lambda_0) + t
o(1)$ be a homotopy defined on $[0,1]\times \overline{B}$, between the homeomorphism
$\Theta(\lambda)-\Theta(\lambda_0)$ and the map $\mathcal{I}_R$.  For sufficiently large
$R$, we have $0 \notin \mathcal{I}_R([0,1]\times \partial B)$.  By degree considerations
\cite{ni:nlfa}, for large $R$, $\mathcal{I}_R$ must hit zero for some $\lambda \in B$.   For
such a $\lambda$, the constraints are satisfied, with the model near infinity corresponding
to $\lambda$ for which $\mathcal{I}_R(\lambda)=0$, and with $\Theta(\lambda)$ near
$\Theta(\lambda_0)$.  \endpf \\

\noindent\textsc{Acknowledgements}. The first author was partially supported by an NSF
Postdoctoral Research Fellowship; the second author acknowledges the partial support of the
NSF through DMS-0104163.

\end{document}